\documentclass{natureinitial}

\usepackage{lineno}
\usepackage{upgreek}
\usepackage{graphicx}
\usepackage{amsmath,amssymb}

\usepackage{multirow}
\usepackage{booktabs}
\usepackage{bm}
\usepackage{makecell}
\usepackage{color}
\usepackage{verbatim}

\title{Pattern of Global Spin Alignment of $\phi$ and $K^{*0}$ mesons in Heavy-Ion Collisions \\ 
}

\author{\rm{STAR Collaboration}}

\begin{document}
\maketitle

\begin{abstract}
Notwithstanding decades of progress since Yukawa first developed a description of the force between nucleons in terms of meson exchange~\cite{Yukawa:1935xg}, 
a full understanding of the strong interaction remains a major challenge in modern science.
One remaining difficulty arises from the non-perturbative nature of the strong
force, which leads to the phenomenon of quark confinement at distances on the order of the size of the proton. Here we show that in relativistic heavy-ion collisions, where quarks and gluons are set free over an extended volume, two species of produced vector (spin-1) mesons, namely $\phi$ and $K^{*0}$, emerge with a surprising pattern of global spin alignment. In particular, the global spin alignment for $\phi$ is unexpectedly large, while that for $K^{*0}$ is consistent with zero. The observed spin-alignment pattern and magnitude for the $\phi$ cannot be explained by conventional mechanisms, while a model with a connection to strong force fields\cite{Sheng:2019kmk,PhysRevD.105.099903,Sheng:2020ghv,Sheng:2022wsy,Sheng:2022ffb}, i.e. an effective proxy description within the Standard Model and Quantum Chromodynamics, accommodates the current data. This connection, if fully established, will open a potential new avenue for studying the behaviour of strong force fields.
\end{abstract}

At the Relativistic Heavy Ion Collider (RHIC) at Brookhaven National Laboratory, heavy ions (e.g., gold nuclei) are accelerated up to 99.995\% of the speed of light and collide from opposite directions. 
Due to the extreme conditions achieved, quarks and gluons are liberated for a brief time ($\sim 10^{-23}$ seconds), instead of being confined inside particles such as protons and neutrons by the strong force. The hot and dense state of matter formed in these collisions is called the quark gluon plasma (QGP)~\cite{Arsene:2004fa,Back:2004je,Adams:2005dq,Adcox:2004mh}. These collisions offer an ideal environment for studying phenomena related to Quantum Chromodynamics, the theory of strong interaction among quarks and gluons. 

In collisions that are not exactly head-on, the approach paths of the two nuclei are displaced by a distance called the impact parameter ($b$), generating a very large orbital angular momentum (OAM) in the system. Part of the OAM is transferred to the QGP in the form of fluid vorticity along the OAM direction which can polarize the spin of the particles through spin-orbit coupling, a phenomenon called global polarization~\cite{Liang:2004ph,Liang:2004xn,Voloshin:2004ha,Betz:2007kg,Becattini:2007sr,Gao:2007bc}. According to the flavour-spin wave function, the polarization of the $\Lambda(\bar{\Lambda}$) hyperon is carried solely by the strange quark $s$ ($\bar{s}$), indicating the global polarization of the $s$ ($\bar{s}$) quark~\cite{Close:1979bt}. The global polarization of $\Lambda(\bar{\Lambda}$) hyperons produced in heavy ion collisions has been studied through their decays by the STAR~\cite{STAR:2017ckg,Adam:2018ivw,STAR:2021beb}, the ALICE~\cite{ALICE:2019onw}, and the HADES~\cite{Kornas:2022cbl} collaborations.

\begin{figure}[h!]       
\centering
\includegraphics[width=0.6 \textwidth]{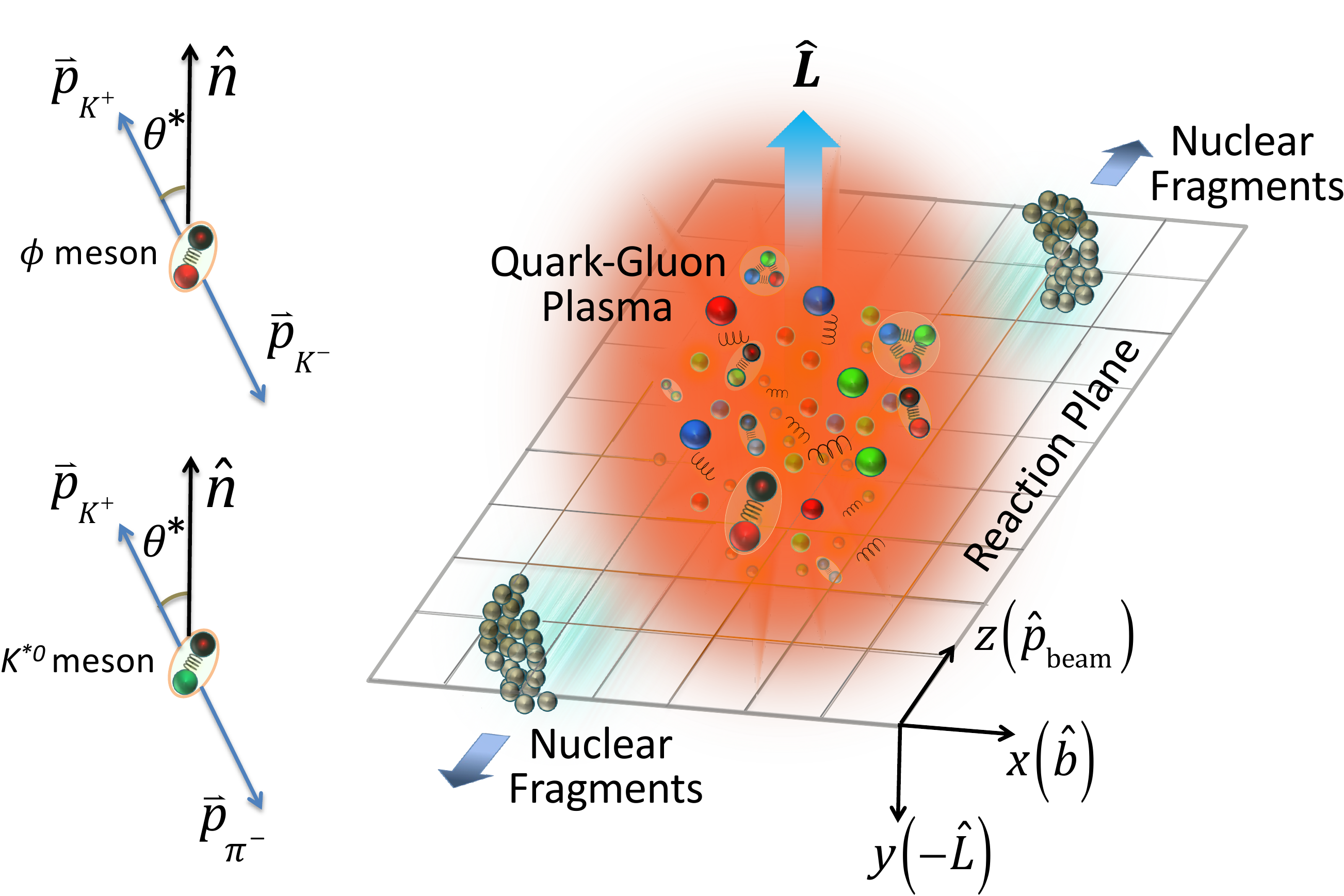}
\caption{\textbf{A schematic view of the coordinate setup for measuring global spin alignment in heavy-ion collisions.} Two nuclei collide and a tiny exploding QGP fireball, only a few femtometers across, is formed in the middle. The direction of the orbital angular momentum ($\hat {L}$) is perpendicular to the reaction plane defined by the incoming nuclei when $b \neq 0$.  The symbol $\vec{p}$ represents the momentum vector of a particle. At the top-left corner, a $\phi$ meson, made of $s$ and $\bar{s}$ quarks, is depicted separately as a particle decaying into a ($K^+$, \,$K^-$) pair. In this example, the quantization axis ($\hat{n}$) for study of the $\phi$ meson's global spin alignment is set to be the same as $\hat{L}$. $\theta^*$ is the polar angle between the quantization axis and the momentum direction of a particle in the decay's rest frame. A similar depiction can be found for a $K^{*0}$ meson at the bottom-left corner.}
  \label{fig:collision}
\end{figure}

The global polarization of quarks influences production of vector mesons such as $\phi(1020)$ and $K^{*0}(892)$. 
Unlike $\Lambda$ ($\bar{\Lambda}$) hyperons, which can undergo weak decay with parity violation, and where the products in the decay's rest frame are emitted preferentially in the spin direction, the polarization of vector mesons cannot be directly measured since they mainly decay through the strong interaction, in which parity is conserved. 
Nevertheless the spin state of a vector meson can be described by a $3\times 3$ spin density matrix with unit trace~\cite{Schilling:1969um}.
The diagonal elements of this matrix, namely, $\rho_{11}, \rho_{00}$ and $\rho_{\rm{-1} \rm{-1}}$, are probabilities for the spin component
along a quantization axis to take the values of 1, 0, and $-$1 respectively. The quantization axis is a chosen axis onto which the projection of angular momentum has well-defined quantum numbers. When the three spin states  have equal probability to be occupied, all three elements are $1/3$ and there is no spin alignment. If $\rho_{00} \neq 1/3$, the probabilities of the three spin states along the quantization axis are different and there is a spin alignment.
In the rest frame of a vector meson decaying to two particles, the angular distribution of one of the decay products can be written as
\begin{eqnarray}\label{eq:extract_rho00}
\frac{dN}{d(\mathrm{cos}\theta^*)} \propto (1-\rho_{00}) + (3 \rho_{00} -1) \mathrm{cos}^2\theta^* ,
\end{eqnarray}
where $\theta^*$ is the polar angle between the quantization axis and the momentum direction of that decay particle. 
By fitting the angular distribution of decay particles with the function above, one can infer the $\rho_{00}$ value.  For the study of global spin alignment, the quantization axis ($\hat{n}$) is chosen to be the direction of the OAM ($\hat {L}$), which is perpendicular to the reaction plane. The reaction plane is defined by the direction of the colliding nuclei (beam direction) and the impact parameter vector ($\hat{b}$)\cite{Poskanzer:1998yz}.  See Fig.~\ref{fig:collision} for a schematic view of the coordinate setup for measuring global spin alignment in heavy-ion collisions. $\phi$ mesons are identified via their decay $\phi \rightarrow K^+ + K^-$. The $K^{*0}$ and $\overline{K^{*0}}$ mesons are reconstructed via their decay $K^{*0} (\overline{K^{*0}}) \rightarrow K^+ \pi^- (K^- \pi^+)$. Hereafter, $K^{*0}$ refers to the combination of $K^{*0}$ and $\overline{K^{*0}}$ unless otherwise specified.

It is assumed~\cite{Sheng:2019kmk,Liang:2004xn,Yang:2017sdk,Xia:2020tyd,Gao:2021rom} that the global spin alignment of $\phi$ mesons can be produced by the coalescence of polarized $s$ and $\bar{s}$ quarks which can be caused by vortical flow or the local fluctuation of mean field (meson field). The conventional sources for the polarization of $s$ and $\bar{s}$ quarks include: the vortical flow~\cite{Becattini:2013vja,Yang:2017sdk} in the QGP in collisions with non-zero impact parameter, the  electromagnetic fields~\cite{Yang:2017sdk,Sheng:2019kmk} generated by the electric currents carried by the colliding nuclei, quark polarization along the direction of its momentum (helicity polarization)~\cite{Gao:2021rom}, and the spin alignment produced by fragmentation of polarized quarks~\cite{Liang:2004xn}. Both the vorticity and electromagnetic fields can be represented as relativistic, rank-2 tensors having ``electric'' (space-time) and ``magnetic'' (space-space) components; each contributes to the quark polarization along the quantization axis $\hat{n}$. For the $\Lambda$ and $\bar{\Lambda}$ polarization in the rest frame, the only contribution is from the magnetic components, in which the vorticity contribution dominates. STAR measurements of the polarization of $\Lambda$ and $\bar{\Lambda}$\cite{STAR:2017ckg,Adam:2018ivw} indicate that the magnetic components of the vorticity and the electromagnetic field tensor in total give~\cite{Sheng:2019kmk,Liang:2004xn,Yang:2017sdk} a negative contribution to $\rho_{00}$ at the level of $10^{-5}$. In addition, the local vorticity loop in the transverse plane~\cite{Xia:2020tyd}, when acting together with coalescence, gives a negative contribution to global $\rho_{00}$. From a hydrodynamic simulation of the vorticity field in heavy-ion collisions, it is known~\cite{Sheng:2019kmk} that the electric component of the vorticity tensor gives a contribution on the order of $10^{-4}$. Simulation of the electromagnetic field in heavy-ion collisions indicates~\cite{Sheng:2019kmk} that the electric field gives a  contribution of order $10^{-5}$.  Fragmentation of polarized quarks contributes on the order of $10^{-5}$, and the effect is mainly present in transverse momenta much larger than a few GeV/$c$.~\cite{Liang:2004xn}. Helicity polarization gives a negative contribution at all centralities~\cite{Gao:2021rom}. Locally fluctuating axial charge currents induced by possible local charge violation gives rise to the expectation~\cite{Muller:2021hpe} of $\rho_{00}(K^{*0}) < \rho_{00}(\phi) < 1/3$.
The aforementioned, mostly conventional mechanisms make either positive or negative contributions to $\phi$ meson $\rho_{00}$, but none of them can produce a $\rho_{00}$ that is larger than 1/3 by more than a few times $10^{-4}$. Recently a theoretical model was proposed based on the $\phi$-meson vector field coupling to $s$ and $\bar{s}$ quarks~\cite{Sheng:2019kmk,PhysRevD.105.099903,Sheng:2020ghv,Sheng:2022wsy,Sheng:2022ffb} analogous to the photon vector field coupled to electrically charged particles. In this mechanism, the observed global spin alignment is caused by the local fluctuation of the strong force field, and can cause deviations of $\rho_{00}$ from 1/3 larger than $10^{-4}$.

In 2008, the STAR collaboration reported on a search for global spin alignment of $\phi(1020)$ and $K^{*0}(892)$ mesons for Au+Au collisions at a center-of-mass energy per nucleon pair of $\sqrt{s_{NN}} = 200$ GeV, with $\hat{n}$ oriented along $\hat{L}$~\cite{Abelev:2008ag}. Due to limited statistics at that time, no significant result was reported. In the present paper we report STAR's measurement of spin alignment for $\phi$ and $K^{*0}$ vector mesons with much larger statistics and at lower collision energies. 

\begin{figure}[h!]       
\centering
\includegraphics[width=0.6 \textwidth]{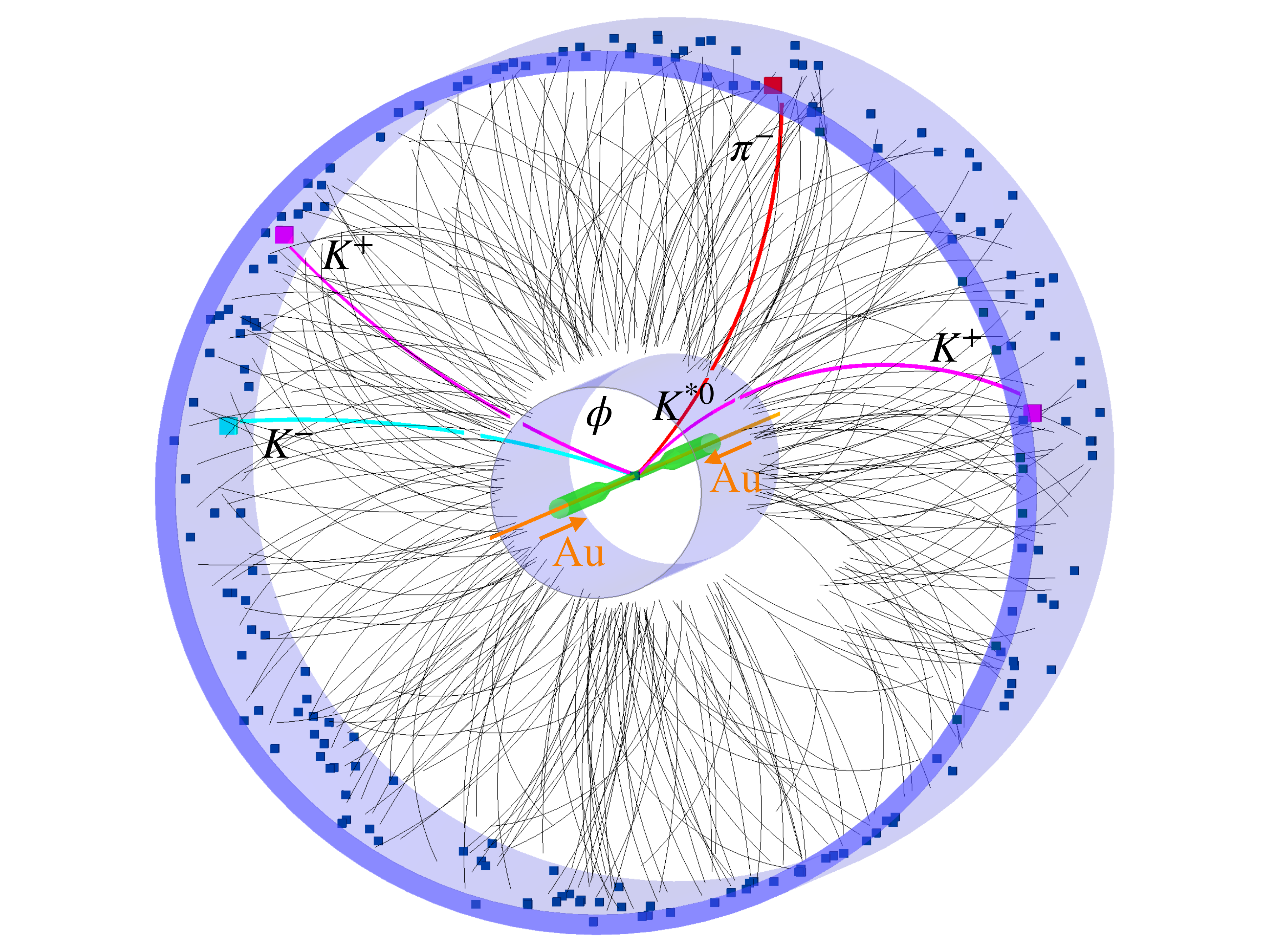}
\caption{\textbf{Schematic display of a single Au+Au collision at $\sqrt{s_{NN}} = 27$ GeV in STAR detector.} A three-dimensional rendering of the STAR TPC, surrounded by the TOF barrel shown as the outermost cylinder. The beam pipe is shown in green and inside it, gold ions travel in opposite directions along the beam axis (brown). Ions collide at the centre of the TPC, and trajectories (gray lines) as well as TOF hits (blue squares) from a typical collision are shown. Reconstructed trajectories of a ($K^+$, $K^-$) pair originating from a $\phi$-meson decay, as well as a $K^+$ and $\pi^-$ from a $K^{*0}$-meson decay, are shown as highlighted tracks.}
  \label{fig:TPC}
\end{figure}

The relevant features of STAR experiment used for the spin alignment measurements are depicted in Fig.~\ref{fig:TPC}.  The two charged daughter particles leave ionization trails inside STAR's Time Projection Chamber (TPC)~\cite{Anderson:2003ur} with trajectories bent in the magnetic field, by which momentum information for charged particles can be reconstructed and the ionization energy loss $(dE/dx)$ inside the gas of the TPC can be calculated. In addition, the time of flight information for particles can be obtained from the Time of Flight (TOF) detector~\cite{Llope:2012zz}, and, combining this with $dE/dx$ measurements, the momentum and particle species for daughters can be determined. Figure~\ref{fig:TPC} shows a three dimensional view of $\phi$ and $K^{*0}$ mesons decaying into their corresponding daughters inside the TPC. More details on the measurement procedure can be found in section Methods.

\begin{figure}[h!]       
\centering
\includegraphics[width=0.6 \textwidth]{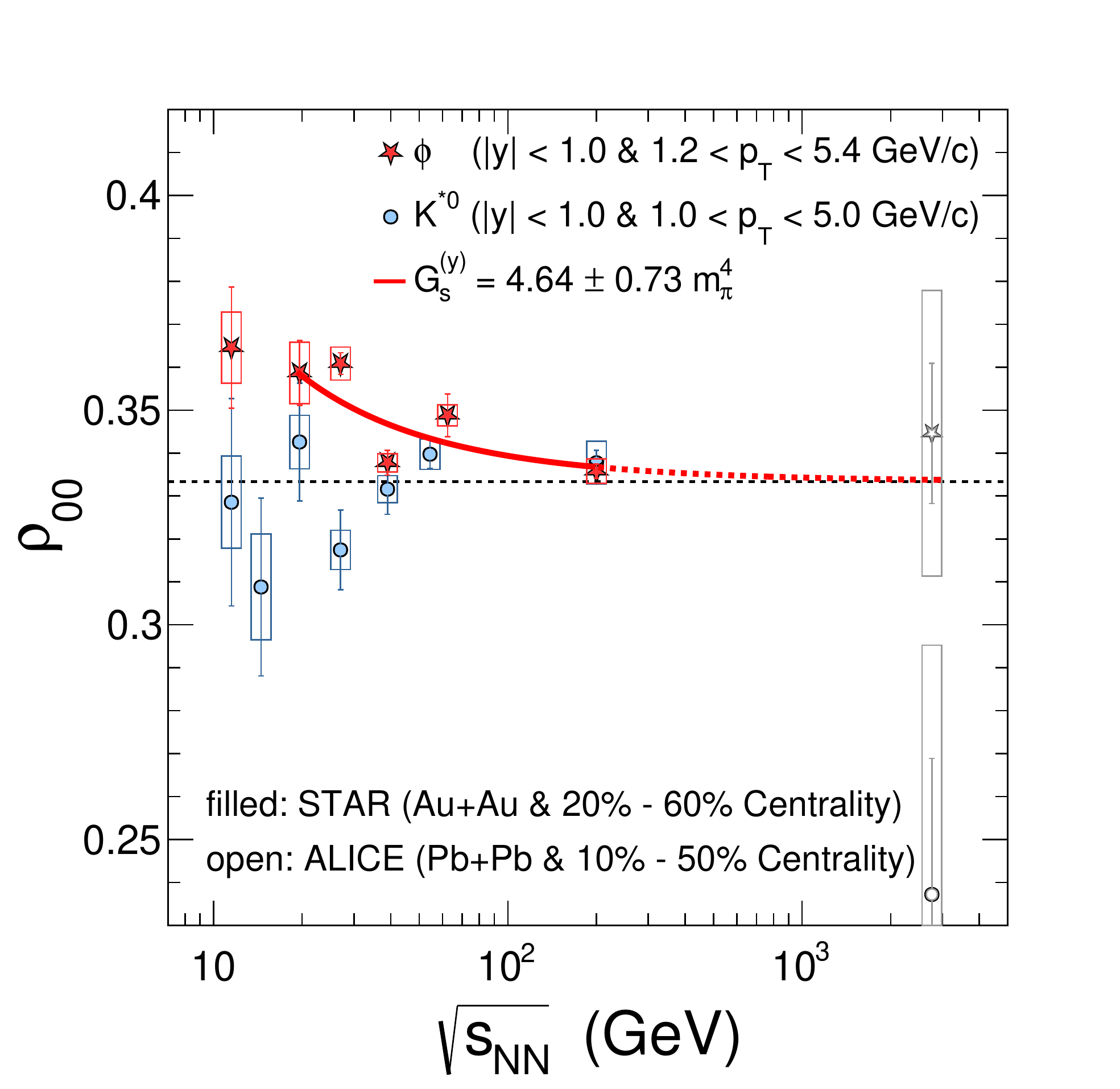}
\caption{\textbf{Global spin alignment of $\phi$ and $K^{*0}$ vector mesons in heavy-ion collisions.} The measured matrix element $\rho_{00}$ as a function of beam energy for the $\phi$ and $K^{*0}$ vector mesons within the indicated windows of centrality, transverse momentum ($p_T$) and rapidity ($y$). The open symbols indicate ALICE results \cite{Acharya:2019vpe} for Pb+Pb collisions at 2.76 TeV at $p_{T}$ values of 2.0 and 1.4 GeV/c for the $\phi$ and $K^{*0}$ mesons, respectively, corresponding to the $p_{T}$ bin nearest to the mean $p_{T}$ for the 1.0 – 5.0 GeV/$c$ range assumed for each meson in the present analysis. The red solid curve is a fit to data in the range of $\sqrt{s_{NN}} = 19.6$ to 200 GeV, based on a theoretical calculation with a $\phi$-meson field \cite{Sheng:2019kmk}. Parameter sensitivity of $\rho_{00}$ to the $\phi$-meson field is shown in Ref.~\cite{Sheng:2022wsy}. The red dashed line is an extension of the solid curve with the fitted parameter  $G_s^{(y)}$. The black dashed line represents $\rho_{00}=1/3.$}
  \label{fig:energyDependence_withTheory}
\end{figure}        

Figure ~\ref{fig:energyDependence_withTheory} shows $\rho_{00}$ for $\phi$ and $K^{*0}$ for Au+Au collisions at beam energies between $\sqrt{s_{NN}} = 11.5$ and 200 GeV.  The centrality categorizes events based on the observed multiplicity of produced charge hadrons emitted from each collision, where 0\% centrality corresponds to exactly head-on collisions, which produce the highest multiplicity, while 100\%
centrality corresponds to barely glancing collisions, which produce the lowest multiplicity.  The STAR measurements presented in Fig.~\ref{fig:energyDependence_withTheory} are for centralities between 20\% and 60\%.  The quantization axis ($\hat{n}$) is taken to be the normal to the 2nd-order event plane~\cite{Poskanzer:1998yz} determined using TPC tracks. The 2nd-order event plane, with its orientation corresponding to the elliptic flow of produced hadrons, serves as a proxy for the reaction plane.
The $\phi$-meson results are presented for $1.2 < p_T < 5.4$ GeV/$c$ and $|y| < 1$. $p_T$ is the momentum in the plane transverse to the beam axis, and rapidity $y = {\rm \tanh^{-1}}\,\beta_z$ with $\beta_z$ being the component of velocity along the beam direction in units of the speed of light.
$\rho_{00}$ for $\phi$-meson is significantly above 1/3 for collision energies of 62 GeV and below, indicating finite global spin alignment. 
The $\rho_{00}$ for $\phi$ mesons, averaged over beam energies between 11.5 and 62 GeV is 0.3512 $\pm$ 0.0017 (stat.) $\pm$ 0.0017 (syst.).
Taking the total uncertainty as the sum in quadrature of statistical and systematic uncertainties, our results indicate that the 
$\phi$-meson $\rho_{00}$ is above 1/3 with a significance of 7.4~$\sigma$.

The $\rho_{00}$ for $K^{*0}$ is shown for $1.0~<~p_T~<~5.0$~GeV/$c$. 
We observe that $\rho_{00}$ for $K^{*0}$ is consistent with 1/3, in marked contrast to the results for $\phi$.
The $\rho_{00}$ for $K^{*0}$, averaged over beam energies of 54.4 GeV and below is 0.3356 $\pm$ 0.0034 (stat.) $\pm$ 0.0043 (syst.).
The complete set of results for $p_T$ and centrality dependence for both vector mesons can be found in the Methods section. Measurements from the ALICE collaboration for Pb+Pb collisions at $\sqrt{s_{\rm NN}}$ = 2.76~TeV~\cite{Acharya:2019vpe}, taken from the closest data points~\cite{Acharya:2019vpe} to the mean $p_T$ for the range of $1.0~<~p_T~<~5.0$~GeV/$c$, are also shown for comparison in Fig. ~\ref{fig:energyDependence_withTheory}.

Intriguingly, $p_T$-averaged $\phi$-meson data at intermediate centrality can be explained by the theoretical model invoking the $\phi$-meson vector field~\cite{Sheng:2019kmk,PhysRevD.105.099903,Sheng:2020ghv,Sheng:2022wsy,Sheng:2022ffb}. This can be seen by fitting the data, as presented by the solid red line in Fig.~\ref{fig:energyDependence_withTheory}. 
This model fit involves adjusting $G_s^{(y)}$ which represents\cite{PhysRevD.105.099903} the quadratic form of field strength tensors multiplied by the effective coupling constant ($g_{\phi}$). In its specific form $G_s^{(y)} \equiv g_{\phi}^2 \big[ 3 \langle B^2_{\phi,y} \rangle + \frac{\langle \textbf{p}^2 \rangle_{\phi}}{m^2_s} \langle E^2_{\phi,y} \rangle - \frac{3}{2} \langle B^2_{\phi,x} + B^2_{\phi,z} \rangle - \frac{\langle \textbf{p}^2 \rangle_{\phi}}{2m^2_s} \langle E^2_{\phi,x} + E^2_{\phi,z} \rangle \big]$, where $E_{\phi,i}$ and $B_{\phi,i}$ are the $i^{\rm th}$-component of the analogous electric and magnetic parts of the $\phi$-meson field, respectively, and $m_s$ is the $s$-quark mass and $\textbf{p}$ its momentum in the $\phi$ rest frame. The stronger deviation of $\rho_{00}$ from 1/3 observed at lower energy is explained by $1/T^2_{\rm eff}$ dependence originating in the theoretical description~\cite{Sheng:2019kmk} from the polarization of quarks in the $\phi$-meson field. Here $T_{\rm eff}$ is the effective temperature of the QGP fireball.  This model can accommodate the large magnitude of $\rho_{00}$ as seen in our measurement, and it also gives the correct collision-energy dependence. The $p_T$ and centrality dependence of the large $\rho_{00}$ signal is recently described by an improved version of the model which is derived from relativistic spin Boltzmann equation~\cite{Sheng:2022wsy}.

The relationship of the $\phi$ meson to the $\phi$-meson field is like that of the photon to the electromagnetic field. 
In analogy to the way in which the photon mediates the electromagnetic interaction, the $\phi$ meson can be regarded as a mediator of the nuclear interaction. 
The $\phi$-meson field behaves like the electromagnetic field since both are vector fields, but the $\phi$-meson field is one component of the short-distance (a few fm) strong force, while the electromagnetic field is a long-distance force. The $\phi$-meson fields, along with other meson fields such as $\sigma$, $\pi$, $\rho$, $\omega$, etc., are low-energy or intermediate-distance (of the order of nuclear radii) 
effective modes of $q\bar{q}$~\cite{Serot:1984ey,Gasser:1983yg}. These modes with the vacuum quantum number are in connection with modes of two gluon fields in Quantum Chromodynamics~\cite{Shifman:1978zn}. 
Just as an electric charge in motion can generate an electromagnetic field, the strange quarks $s$ and $\bar{s}$ in motion can produce an effective $\phi$-meson field. The local difference between the currents of $s$ and $\bar{s}$ (net-strangeness current), which may occur because the $s$ and $\bar{s}$ have different momenta at a given space-time point, can generate an effective $\phi$-meson field. Through its magnetic part, the vector meson field has been used to predict the difference between the polarization of $\Lambda$ and $\bar{\Lambda}$~\cite{Csernai:2018yok}. Similar to how an electric field can polarize a quark and anti-quark through spin-orbit couplings, the strong electric part of the $\phi$-meson field can also polarize $s$ and $\bar{s}$, leading to a positive contribution to $\rho_{00}$ of the $\phi$-meson (as a bound state of $s$ and $\bar{s}$) but with much larger magnitude due to its strong interaction (a large coupling constant $g_\phi$). Figure~\ref{fig:energyDependence_withTheory} shows that, while conventional explanations fall far short in accounting for the data, our experimental measurement in 20-60\% centrality can be described well by this model, which invokes the $\phi$-meson field, thus favoring the conclusion that the $\phi$-meson field leads to the $\phi$-meson global spin alignment.

The lifetime of $K^{*0}$ is about 10 times shorter than the $\phi$ lifetime, corresponding to a mean proper decay length $c\tau \approx 4.1$~fm, making it susceptible to in-medium effects. The difference between the global spin alignment for $K^{*0}$ and $\phi$ may be attributed to different in-medium interactions due to this difference in lifetime, a polarization transfer during the late stage of hadronic interactions~\cite{Karpenko:2016jyx}, and a different response to the vector meson field~\cite{Sheng:2019kmk}. Similar to strange quarks ($s$ and $\bar{s}$), light quarks can also be polarized by vorticity fields and vector meson fields. 
However, the vector fields that polarize light quarks, such as the $\rho$ and $\omega$ fields, are distinct from the $\phi$ field that polarizes strange quarks. The contributions from vector meson fields to $\rho_{00}$ for $K^{*0}$ involve averages of products of different vector meson fields such as that from the $\phi$ (for the $\bar{s}$) and $\rho$ (for the $d$). It is expected that the correlations between these two different, fluctuating vector meson fields for $d$ and $\bar{s}$ are much weaker than the correlations between the same fields for $s$ and $\bar{s}$, causing the vector meson field contributions to $\rho_{00}$ for $K^{*0}$ to be negligible \cite{Sheng:2020ghv}. The above considerations may account for the insignificant deviation of $\rho_{00}$ for $K^{*0}$ from 1/3,
as observed in experiments. A comprehensive and quantitative study of all these effects is needed to reveal the nature of such a significant difference between spin alignments of $K^{*0}$ and $\phi$. Our new data provides motivation for further theoretical developments in this direction.

Based on the fit to our data in Fig.~\ref{fig:energyDependence_withTheory} with the model in Ref.~\cite{Sheng:2019kmk}, we estimate the free parameter in the fit, $G_s^{(y)}$, to be $4.64 \pm 0.73 m^4_{\pi}$. This value of $G_s^{(y)}$ is compatible with the value of the average field squared times $g_\phi^2$ used in the calculation of relativistic Spin Boltzmann equation~\cite{Sheng:2022wsy}.
The extracted value serves as only a rough estimate, as uncertainties and assumptions in Ref.~\cite{Sheng:2019kmk} await further studies by the theoretical community. This is a qualitatively new class of measurement, and it offers important guidance for future theoretical progress concerning the strong force field under extreme conditions.

Measurements of the global spin alignment of vector mesons provide new knowledge about the vector meson fields. The vector meson fields are an essential part of the nuclear force that binds nucleons inside atomic nuclei \cite{Bryan:1969mp,Nagels:1977ze} and are also pivotal in describing properties of nuclear structure and nuclear matter \cite{Walecka:1974qa,Serot:1984ey}. The $\rho_{00}$ for the $\phi$ meson has a desirable feature in that all contributions depend on squares of field amplitudes; it can be regarded as a field analyzer \cite{Sheng:2019kmk}  which makes it possible to extract the imprint of the $\phi$-meson field even if the field fluctuates strongly in space-time. Another important feature worthy of mention is that an essential contribution to the $\phi$-meson $\rho_{00}$ is from the term~\cite{Sheng:2019kmk} $\sim \mathbf{S}\cdot (\mathbf{E}_\phi \times \mathbf{p})$, where $\mathbf{E}_\phi$ is the electric part of the $\phi$-meson field induced by the local, net strangeness current density, and $\mathbf{S}$ and $\mathbf{p}$ are the spin and momentum of the strange (anti)quarks, respectively. Such a term is nothing but the quark version of the spin-orbit force which, at the nucleon level, plays a key role in the nuclear shell structure \cite{Mayer:1949pd,Haxel:1949fjd}. Our measurements of a signal based on global spin alignment for vector mesons reveal a surprising pattern and a value for $\phi$ meson that is orders of magnitude larger than can be explained by conventional effects. This work provides a potential new avenue for understanding the strong interaction at work at the sub-nucleon level.


\clearpage

\setcounter{figure}{0}
\setcounter{equation}{0}
\renewcommand{\figurename}{Extended\ Data\ Figure}
\renewcommand{\tablename}{Extended\ Data\ Table}

\begin{figure}[h!]       
\centering
\includegraphics[width=0.7\textwidth]{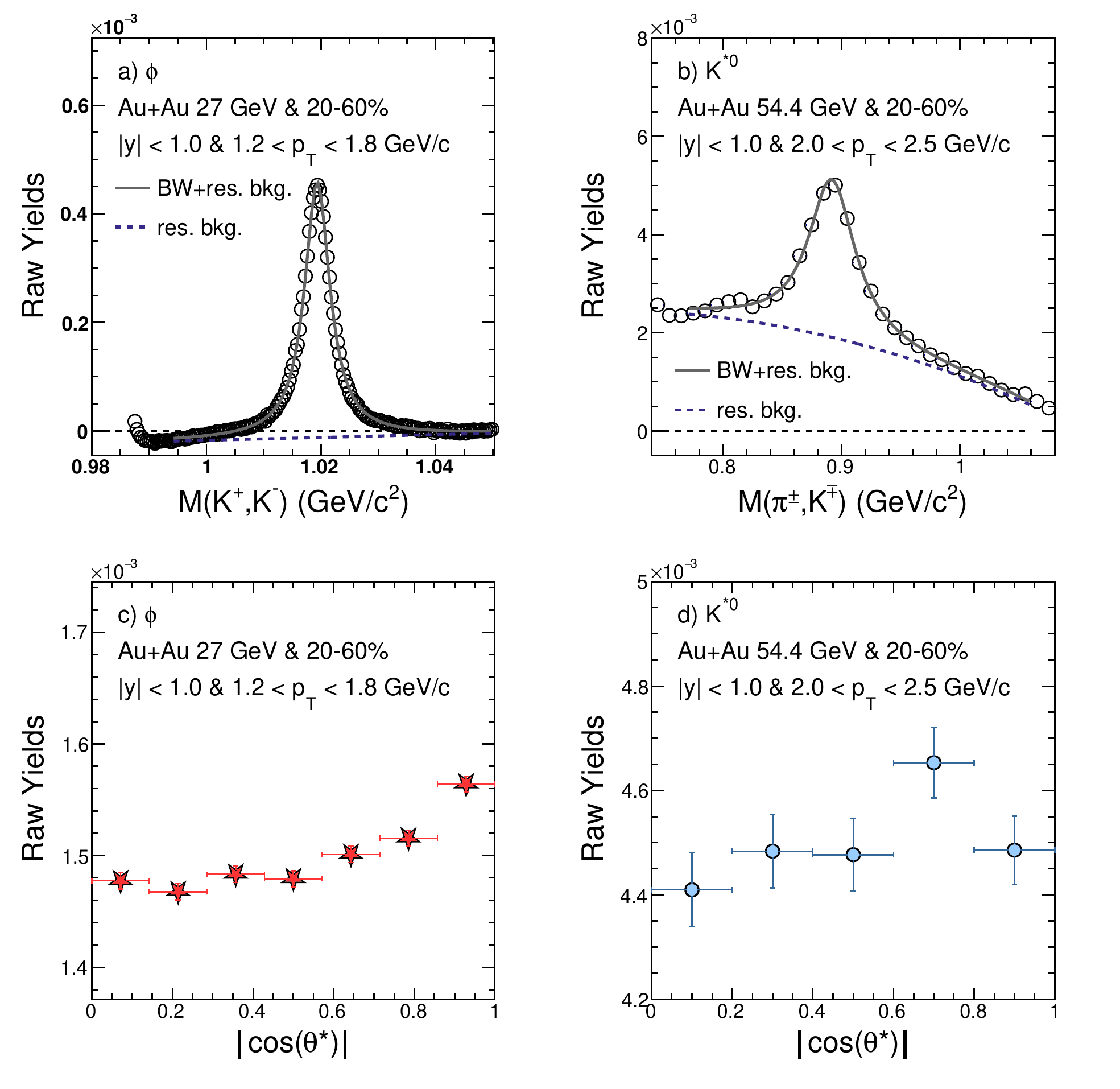}
\caption{\textbf{Example of combinatorial background subtracted invariant mass distributions and the extracted yields as a function of $\cos \theta^*$ for $\phi$ and $K^{*0}$ mesons.} \textbf{a)} example of $\phi \rightarrow K^+ + K^-$ invariant mass distributions, with combinatorial background subtracted, integrated over $\cos \theta^*$; \textbf{b)} example of $K^{*0} (\overline{K^{*0}}) \rightarrow K^{-} \pi^{+} (K^{+} \pi^{-})$ invariant mass distributions, with combinatorial background subtracted,  integrated over $\cos \theta^*$; \textbf{c)} extracted yields of $\phi$ as a function of $\cos \theta^*$; \textbf{d)} extracted yields of $K^{*0}$ as a function of $\cos \theta^*$.}
\label{fig:imass}
\end{figure}

\begin{methods}

\subsection{Data description} 
This $\phi$-meson $\rho_{00}$ analysis is based on Au+Au collisions at $\sqrt{s_{\rm NN}}$ = 11.5, 19.6, 27, 39, 62.4, and 200\,GeV, with samples of 8, 19, 348, 117, 45, and 1560 million events, respectively. For $K^{*0}$ mesons, the sample sizes are 12, 18, 36, 70, 130, 520, and 350 million events at $\sqrt{s_{\rm NN}}$ = 11.5, 14.5, 19.6, 27, 39, 54.4, and 200\,GeV, respectively. All data were taken using a minimum-bias trigger (MB). This trigger selects all particle-producing collisions regardless of the extent of overlap of the incident nuclei. To maximize the statistics and ensure uniform acceptance, a selection on the position of the reconstructed primary vertex along the beam axis ($V_z$) is made for each of the energies. In the case of the $\phi$ analysis, $V_z$ is required to be within $\pm$30 cm of the centre of the STAR Time Projection Chamber~\cite{Anderson:2003ur} for $\sqrt{s_{NN}} = 200$ GeV, while the corresponding $V_z$ windows are $\pm 40$, 40, 70, 70, and 50 cm at beam energies of 62.4, 39, 27, 19.6, and 11.5 GeV, respectively. For $K^{*0}$, the $V_z$ window is $\pm 50$ cm at 39 GeV and below, and $\pm 30$ cm at the remaining beam energies.  
Charged particles with pseudo-rapidities $|\eta| < 1.0$ are reconstructed using the TPC.  
For both analyses, the centrality definition is based on the raw charged particle multiplicity in the TPC within  $|\eta| < 0.5$.  
The primary vertex position in the plane that is transverse to the direction of the colliding Au ion beams, $V_r$, is required to be within 2 cm of the peak of the reconstructed primary vertex position for all energies except 14.5 GeV.  For 14.5 GeV the vertex is not centred at (0, 0) in the x-y plane and slightly offset at (0.0, $-$0.89) cm, and the $|V_r|$ (=$\sqrt{V_{x}^{2} + (V_{y}+0.89)^{2}}$) is selected to be smaller than 1 cm to reject interactions with the beam pipe.

\subsection{Reconstruction of event plane} 
In this paper, we follow the same procedure as in STAR's previous study~\cite{Abelev:2008ag}, by using the 2nd-order event plane (EP) based on tracks in the TPC as a proxy for the event reaction plane. $\phi$ and $K^{*0}$ daughter candidates were excluded from the event plane determination, to avoid self-correlation between EP and those particles under study. In addition, results obtained using the 1st-order EP are presented in this section for the $\phi$ global spin alignment. The 1st-order EP is based on the  Shower Maximum Detectors (SMD) of the Zero Degree Calorimeters (ZDC)~\cite{Adler:2000bd} for the $\sqrt{s_{NN}} = 62.4$ and 200 GeV data, and on the Beam-Beam Counter \cite{Whitten:2008zz, Allgower:2002zy} for the lower energies. 

In non-central collisions, a fraction of the initial angular momentum is carried away by spectator nucleons, and therefore the normal to the 1st-order EP can be more sensitive to the direction of the initial global angular momentum than that for the 2nd-order EP. On the other hand, the resolution of the 2nd-order EP, based on the TPC tracking, is better than that of the 1st-order EP, owing to the large multiplicity and elliptic flow~\cite{Poskanzer:1998yz} within the TPC acceptance near middle rapidity. As discussed in Ref. \cite{Tang:2018qtu}, when all corrections are taken into account, the two measurements should agree with each other to first-approximation, as demonstrated below. Uncertainties in the event plane resolution are negligible relative to the statistical and systematic uncertainties of the final results.

\subsection{$\phi$- and $K^{*0}$-meson yield extraction} 
The distributions of $\phi$ and $K^{*0}$ invariant mass are obtained for each $p_T$, centrality, and $\cos\theta^*$ bin. The corresponding combinatorial background for the $\phi$ meson is estimated by event mixing, i.e., creating $K^+,\,K^-$ pairs from tracks selected from different events with the same centrality, event plane angle bin, and primary vertex bin. For $K^{*0}$ mesons, the background is estimated by rotating the momentum vector of one of the decay daughters by $180^\circ$. Both techniques can effectively break the correlation between pairs in real events, and the results from the two techniques are consistent within $1.0 \sim 1.5 \sigma$. Invariant mass yields are then obtained by subtracting the corresponding backgrounds. Small, residual backgrounds remain, due to particle misidentification for both techniques, and to non-resonance correlations for the rotation technique. The upper panels of Extended Data Fig.~\ref{fig:imass} show typical combinatorial background subtracted $\phi$ and $K^{*0}$ invariant mass distributions integrated over $\cos \theta^*$. The extracted yield is fitted with a Breit-Wigner function for the signal, plus a second-order polynomial curve for the residual background. The lower panels of Extended Data Fig.~\ref{fig:imass} show examples of $\phi$ and $K^{*0}$ yield as a function of $\cos\theta^*$. This yield, after correction for detection efficiency and acceptance at each $p_T$ and centrality, is then used to extract $\rho_{00}$.

\subsection{Corrections for finite EP resolution, efficiency, and acceptance \\}

 i) {\bf $\phi$-meson $\rho_{00}$ analysis}  ~~Detector efficiency within the acceptance is corrected using the STAR Monte Carlo embedding method \cite{Abelev:2009bw, Adams:2004ep, Aggarwal:2010mt}. To account for finite EP resolution and finite acceptance in pseudo-rapidity ($\eta$)~\cite{Lan:2017nye}, the observed $\cos\theta^*$ distribution is not fitted using Eq. 1 in the main text, but is instead described by the correction procedure derived in Ref.~\cite{Tang:2018qtu} wherein the data are fitted using
\begin{eqnarray}
\begin{aligned}
\left[ \frac{dN}{d\cos\theta^*} \right]_{|\eta|}
\propto &  (1 + \frac{B'F}{2}) + (A'+F)\cos^2{\theta^*} \\
+ & (A'F-\frac{B'F}{2})\cos^4{\theta^*},
\end{aligned}
\label{eq:fitEqEffXAccpt_00}
\end{eqnarray}
where 
\begin{eqnarray}
A'=\frac{A(1+3R)}{4+A(1-R)} , ~~~
B'=\frac{A(1-R)}{4+A(1-R)}  ,
\end{eqnarray}
 and 
\begin{eqnarray}
A=\frac{3\rho_{00}-1}{1-\rho_{00}} ,
\label{eq:Arho_00}
\end{eqnarray}
and $F$ is a factor that accounts for finite acceptance. Its value depends on $p_T$ and $\eta$ and is calculated using a simulation~\cite{Tang:2018qtu}. The factor $R$ accounts for finite EP resolution. For the 1st-order EP, it is $R_1=\langle \mathrm{cos}2(\Psi_r-\Psi_1) \rangle$, where $\Psi_1$ is the first order EP and $\Psi_r$ is the true reaction plane. $R_1$ can be obtained following the usual procedure in flow analyses~\cite{Poskanzer:1998yz}. For the 2nd-order EP, $R$ is replaced by $R_{21}=\langle{\mathrm{cos}2( \Psi_1 - \Psi_2 )}\rangle / R_1$, where $\Psi_2$ is the 2nd-order EP. Extended Data Fig.~\ref{fig:rhoSigCorrPhi} shows an example of such fitting. The fitting procedure has been repeated with different $\eta$ acceptance cuts for the decay daughters, namely $|\eta|<1$ and $|\eta|<0.6$, and results after correction converge as expected, as seen in simulations~\cite{Tang:2018qtu}. In this procedure, the corrections for detector efficiency and acceptance are applied separately. Doing it this way provides insight into the effect of acceptance alone, and the effect of acceptance can be taken into account with a high precision. In practice, this procedure has been verified to give results consistent with those from procedure ii) below. It is worth noting that, in simulation studies, we found that the decay topology dependent efficiency along with the elliptic flow ($v_{2}$) ~\cite{Poskanzer:1998yz} of the parent meson can bias the $\rho_{00}$ measurements. Such effects have been fully corrected with the procedure of efficiency correction, for both $\phi$ and $K^{*0}$.

\noindent ii) {\bf $K^{*0}$ $\rho_{00}$ analysis} ~~The detector acceptance and efficiency are calculated using the STAR Monte Carlo embedding method \cite{Abelev:2009bw, Adams:2004ep, Aggarwal:2010mt}.  In this process, a small additional fraction of $K^{*0}$ mesons (5\%) is generated with a uniform distribution in the rapidity range [$-$1,1], transverse momentum range [0, 10 GeV/$c$], and azimuthal angle range [0, 2$\pi$], and then passed through the STAR detector simulation in GEANT3~\cite{Fine:2000qx}. The number of $K^{*0}$ mesons reconstructed after passing through the detector simulation and through the same set of track selections as used in real data, compared to the input number of $K^{*0}$ within the same rapidity interval, gives the reconstruction efficiency $\times$ acceptance ($\epsilon_{\rm rec}$).
The yield, after the correction for reconstruction efficiency $\times$ acceptance, is fitted with 
\begin{eqnarray}
\frac{dN}{d(\cos\theta^*)} \propto (1-\rho^{\rm obs}_{00}) + (3 \rho^{\rm obs}_{00} -1) \cos^2\theta^* 
\label{eq:abservedRho_00}
\end{eqnarray}
to extract $\rho^{\rm obs}_{00}$, where "obs" stands for "observed". Extended Data Fig.~\ref{fig:rhoSigCorrK} shows an example of such fitting. The $\rho^{\rm obs}_{00}$ is then corrected for finite EP resolution ($R$), following the procedure laid out in Ref~\cite{Tang:2018qtu}, to obtain the final $\rho_{00}$,
\begin{eqnarray}
\rho_{00} - \frac{1}{3} = \frac{4}{1+3R}(\rho^{\rm obs}_{00} - \frac{1}{3}).
\end{eqnarray}
The stability of the embedding correction is validated by repeating the analysis with the procedure in i), and both procedures give consistent results.

\subsection{Consistency check using the 1st-order event plane}
In Extended Data Fig.~\ref{fig:energyDependence_with1stEP}, the $\rho_{00}$ of $\phi$ mesons at $p_T > 1.2$ GeV/$c$ is presented for Au+Au collisions at $\sqrt{s_{NN}} = 11.5, 19.6, 27, 39, 62.4$, and 200 GeV. For $1.2 < p_T < 5.4$ GeV/$c$, $\rho_{00}$ averaged over energies of 62.4 GeV and below is 0.3565 $\pm$ 0.0037 (stat.) $\pm$ 0.0042 (syst.) for the 1st-order EP, and 0.3512 $\pm$ 0.0017 (stat.) $\pm$ 0.0017 (syst.) for the 2nd-order EP. The former has a larger error than the latter, due to its lower EP resolution. Taking the total uncertainty as the quadrature sum of statistical and systematic errors, the two measurements are consistent with each other within $\sim 2 \sigma$. Both measurements indicate strong global spin alignment with a 4.2 $\sigma$ (1st-order EP) and 7.4 $\sigma$ (2nd-order EP) significance.
For $K^{*0}$, the 1st-order EP result is not presented since the statistical errors are too large due to the lower 1st-order EP resolution.

\subsection{Self-consistency check with randomly oriented $\hat{L}$}
As a self-consistency check for the procedure, we also repeated both analyses with the $\hat{L}$ direction randomly oriented in space, for which any global spin alignment would be eliminated and $\rho_{00}$ should be 1/3. Our exercise with randomly oriented $\hat{L}$ gives 0.3378 $\pm$ 0.0016 (stat.) $\pm$ 0.0010 (syst.) for the $\phi$ meson and 0.3369 $\pm$ 0.0086 (stat.) $\pm$ 0.0053 (syst.) for $K^{*0}$ (averaged over beam energies of 62.4 GeV and below). 

\subsection{Global spin alignment in the in-plane direction} 
Extended Data Fig.~\ref{fig:energyDependence_withInPlane} shows $\rho_{00}$ for $\phi$ with two choices of quantization axes that are perpendicular to each other, namely, $\hat{L}$ and $\hat{b}$, corresponding to the out-of-plane and in-plane directions, respectively. $\hat{L}$ is the usual choice of quantization axis and is used everywhere else in this paper. Note that although the direction of $\hat{b}$ is rotated $90^\circ$ about the beam axis relative to $\hat{L}$ in the ideal case, the $\theta^*$ angles obtained with $\hat{b}$ and $\hat{L}$ do not differ by $90^\circ$ in general. Thus their $\rho_{00}$ cannot be mapped to each other by a trivial relationship. The plot shows that $\rho_{00}$ in the out-of-plane direction is considerably larger than in the in-plane direction, which can be attributed to the effect of elliptic flow~\cite{Sheng:2022wsy}.

\subsection{Transverse momentum dependence} 
Extended Data Fig.~\ref{fig:ptDependencePhi} and Fig.~\ref{fig:ptDependenceKstar} show $\rho_{00}$ as a function of transverse momentum for $\phi$ and $K^{*0}$, respectively.  At low transverse momentum ($150 < p_T < 400$ MeV/$c$) the TPC tracking efficiency rises steeply with increasing $p_{T}$, and consequently, there is a bias against a daughter kaon pairing with another kaon from the adjacent phase space. This constraint in forming pairs introduces a significant artificial $\phi$-meson $\rho_{00}$ at relatively low $p_T$ that is difficult to correct.  For that reason, $\rho_{00}$ for $\phi$ mesons is presented for $p_T > 1.2$ GeV/$c$ only, where the aforementioned effect diminishes and measurements are reliable, as confirmed by simulation studies. For a similar reason, $\rho_{00}$ for $K^{*0}$ is shown for $p_T > 1.0$ GeV/$c$ only. For all energies considered, we see that the departure of $\rho_{00}$ from 1/3 for the $\phi$ meson occurs mainly at $p_T$ within $\sim$ 1.0 - 2.4 GeV/$c$, and at larger $p_T$ the result can be regarded as being consistent with 1/3 within $\sim 2\sigma$ or less. The measurement of energy and centrality dependence shown in this paper were obtained by averaging $\rho_{00}$($p_T$) discussed in this subsection for the corresponding centrality and $p_T$ range with $1/(\mathrm{stat. \, error})^2$ as weight. We compared the $\rho_{00}$ value for $\phi$ meson at 27 GeV (our best statistical data point) to the yield weighted average and the difference is negligible.

\subsection{Centrality dependence} 
Extended Data Fig.~\ref{fig:centDependence} shows $\rho_{00}$ as a function of centrality at selected energies, for $\phi$ (upper panels) and $K^{*0}$ (lower panels). The $p_T$ range for taking the average value for $\phi$ is $1.2 < p_T < 5.4$ GeV/$c$ and for $K^{*0}$ is $1.0 < p_T < 5.0$ GeV/$c$. At high energies (62.4 GeV and above for $\phi$ mesons, 39 GeV and above for $K^{*0}$), $\rho_{00}$ in central collisions tends to be less than 1/3. This might be caused by transverse local spin alignment~\cite{Xia:2020tyd} and/or a contribution from the helicity polarization of quarks~\cite{Gao:2021rom} which tend to reduce $\rho_{00}$. This reduction in central collisions is further examined by plotting $\rho_{00}$ as a function of energy for central collisions, as shown in Extended Data Fig.~\ref{fig:energyDependenceCentral}. We see that the  $\rho_{00}$ of $\phi$ mesons for 0-20\% central collisions decreases with increasing energy, and deviates below 1/3 with marginal significance at $\sqrt{s_{\rm NN}}$ = 200 GeV. The $p_T$ and centrality dependence of the large $\rho_{00}$ signal is recently described by an improved version of the model with $\phi$-meson field~\cite{Sheng:2022wsy}. In this updated work, instead of considering a static meson, the global spin alignment is first derived from spin Boltzmann Equation in $\phi$-meson's rest frame, and then transformed into laboratory frame with known momentum.

\subsection{Global and local spin alignment} 
In heavy ion collisions, the global spin alignment for a collision system can show up in local spin alignments also.  It is the same phenomenon, but viewed from different frames. For example, the relation between global $\rho_{00}$ and production plane $\rho_{00}\{\mathrm{PP}\}$ is given~\cite{Acharya:2019vpe} by $\rho_{00}\{\mathrm{PP}\} - \frac{1}{3} = (\rho_{00} - \frac{1}{3}) \frac{1+3v_2}{4} $. Here the production plane is the plane defined by the beam and vector meson's momentum direction, and the $\rho_{00}\{\mathrm{PP}\}$ is measured with the normal to the production plane as the quantization axis. Another popular choice of local frame is the helicity frame, in which vector meson's momentum direction is taken as the quantization axis. An analytical relation between global $\rho_{00}$ and the helicity frame $\rho_{00}$ does not exist, but based on our simulation for the same kinematic range, typical values of $\rho_{00}$ in the helicity frame (in between 0.2 and 0.6~\cite{Liang:2020}) will result in the global $\rho_{00}$ deviating from 1/3 by only $\sim 0.001$ and $\sim 0.01$ for $\phi$ and $K^{*0}$ mesons, respectively, which are either negligible or very small when compared to the ($\rho_{00}-1/3$) observations presented in this work. In a recent work, it is argued that the gradient of the radial flow along the beam axis can generate transverse vorticity loops at finite rapidity, and cause the transverse local spin alignment~\cite{Xia:2020tyd}. This effect can give a negative contribution to the global spin alignment of vector mesons, and is more prominent and clearly evident in central collisions. This can be part of the reason why at top RHIC energies, we observe that the central value of $\rho_{00}$ is below 1/3. 

\subsection{The result with $\hat{L}$ boosted into vector meson's rest frame}
In the study of the hyperon global polarization or the vector meson global spin alignment, it is a convention to take $\hat{L}$ in the laboratory frame as the quantization axis. We follow that convention in this paper. An alternative choice of the quantization axis is the direction of $\hat{L}$ after being boosted into particle's rest frame~\cite{Florkowski:2021pkp}. We estimated that for our $\rho_{00}$ value that is averaged over beam energies of 62.4 GeV and below, the difference between the results with and without boosting $\hat{L}$ into rest frame is on the order of $10^{-3}$.

\subsection{Taking the average value of 62.4 GeV and below}
For both $\phi$ and $K^{*0}$, the averaged $\rho_{00}$ value of 62.4 GeV and below is obtained by taking the average with $1/(\mathrm{stat. \, error})^2$ as weight for each energy. 

\subsection{Systematic error}
For each beam energy, sources of systematic uncertainty can be categorized as i) quality selections at the event and track level, ii) particle identification cuts, iii) several invariant mass fitting ranges and residual background functions (first- and second-order polynomials) for signal extraction, iv) histogram bin counting vs. functional integration for yield extraction, v) different efficiency evaluation methods.
After repeating the analysis with reasonable variations of quality selections or analysis procedures and obtaining the corresponding values, systematic errors from each individual source are calculated as $(\mathrm{maximum\, value - minimum\, value})/\sqrt{12}$, assuming uniform probability distributions between the maximum and minimum values. The final systematic errors are the quadrature sum of the systematic errors from the various sources.
The averaged $\rho_{00}$ over beam energies of 62.4 GeV and below is calculated for each variation. The systematic errors for averaged $\rho_{00}$ are evaluated with the same procedure as described above. 
Contributions of each systematic uncertainty for the averaged $\rho_{00}$ are listed in extended data Tables ~\ref{table:phiSysError} and ~\ref{table:kStarSysError}, for $\phi$ and  $K^{*0}$ respectively.

\end{methods}

\bibliographystyle{unsrt}
\bibliography{reference}

\begin{thebibliography}{10}

\bibitem{Yukawa:1935xg}
Hideki Yukawa.
\newblock {On the Interaction of Elementary Particles I}.
\newblock {\em Proc. Phys. Math. Soc. Jap.}, 17:48--57, 1935.

\bibitem{Sheng:2019kmk}
Xin-Li Sheng, Lucia Oliva, and Qun Wang.
\newblock {What can we learn from the global spin alignment of $\phi$ mesons in
  heavy-ion collisions?}
\newblock {\em Phys. Rev. D}, 101(9):096005, 2020.

\bibitem{PhysRevD.105.099903}
Xin-Li Sheng, Lucia Oliva, and Qun Wang.
\newblock Erratum: What can we learn from the global spin alignment of
  $\ensuremath{\phi}$ mesons in heavy-ion collisions? [phys. rev. d 101, 096005
  (2020)].
\newblock {\em Phys. Rev. D}, 105:099903, May 2022.

\bibitem{Sheng:2020ghv}
Xin-Li Sheng, Qun Wang, and Xin-Nian Wang.
\newblock {Improved quark coalescence model for spin alignment and polarization
  of hadrons}.
\newblock {\em Phys. Rev. D}, 102(5):056013, 2020.

\bibitem{Sheng:2022wsy}
Xin-Li Sheng, Lucia Oliva, Zuo-Tang Liang, Qun Wang, and Xin-Nian Wang.
\newblock {Spin alignment of vector mesons in heavy-ion collisions}.
\newblock 5 2022.

\bibitem{Sheng:2022ffb}
Xin-Li Sheng, Lucia Oliva, Zuo-Tang Liang, Qun Wang, and Xin-Nian Wang.
\newblock {Relativistic spin dynamics for vector mesons}.
\newblock 6 2022.

\bibitem{Arsene:2004fa}
I.~Arsene et~al.
\newblock {Quark gluon plasma and color glass condensate at RHIC? The
  Perspective from the BRAHMS experiment}.
\newblock {\em Nucl. Phys.}, A757:1--27, 2005.

\bibitem{Back:2004je}
B.~B. Back et~al.
\newblock {The PHOBOS perspective on discoveries at RHIC}.
\newblock {\em Nucl. Phys.}, A757:28--101, 2005.

\bibitem{Adams:2005dq}
John Adams et~al.
\newblock {Experimental and theoretical challenges in the search for the quark
  gluon plasma: The STAR Collaboration's critical assessment of the evidence
  from RHIC collisions}.
\newblock {\em Nucl. Phys.}, A757:102--183, 2005.

\bibitem{Adcox:2004mh}
K.~Adcox et~al.
\newblock {Formation of dense partonic matter in relativistic nucleus-nucleus
  collisions at RHIC: Experimental evaluation by the PHENIX collaboration}.
\newblock {\em Nucl. Phys.}, A757:184--283, 2005.

\bibitem{Liang:2004ph}
Zuo-Tang Liang and Xin-Nian Wang.
\newblock {Globally polarized quark-gluon plasma in non-central A+A
  collisions}.
\newblock {\em Phys. Rev. Lett.}, 94:102301, 2005.
\newblock [Erratum: Phys. Rev. Lett.96,039901(2006)].

\bibitem{Liang:2004xn}
Zuo-Tang Liang and Xin-Nian Wang.
\newblock {Spin alignment of vector mesons in non-central A+A collisions}.
\newblock {\em Phys. Lett. B}, 629:20--26, 2005.

\bibitem{Voloshin:2004ha}
Sergei~A. Voloshin.
\newblock {Polarized secondary particles in unpolarized high energy
  hadron-hadron collisions?}
\newblock 2004.

\bibitem{Betz:2007kg}
Barbara Betz, Miklos Gyulassy, and Giorgio Torrieri.
\newblock {Polarization probes of vorticity in heavy ion collisions}.
\newblock {\em Phys. Rev. C}, 76:044901, 2007.

\bibitem{Becattini:2007sr}
F.~Becattini, F.~Piccinini, and J.~Rizzo.
\newblock {Angular momentum conservation in heavy ion collisions at very high
  energy}.
\newblock {\em Phys. Rev. C}, 77:024906, 2008.

\bibitem{Gao:2007bc}
Jian-Hua Gao, Shou-Wan Chen, Wei-tian Deng, Zuo-Tang Liang, Qun Wang, and
  Xin-Nian Wang.
\newblock {Global quark polarization in non-central A+A collisions}.
\newblock {\em Phys. Rev. C}, 77:044902, 2008.

\bibitem{Close:1979bt}
F.~E. Close.
\newblock {\em {An Introduction to Quarks and Partons}}.
\newblock Academic Press Inc. (London) 1979, 481p, 1979.

\bibitem{STAR:2017ckg}
L.~Adamczyk et~al.
\newblock {Global $\Lambda$ hyperon polarization in nuclear collisions:
  evidence for the most vortical fluid}.
\newblock {\em Nature}, 548:62--65, 2017.

\bibitem{Adam:2018ivw}
Jaroslav Adam et~al.
\newblock {Global polarization of $\Lambda$ hyperons in Au+Au collisions at
  $\sqrt{s_{NN}}$ = 200 GeV}.
\newblock {\em Phys. Rev. C}, 98:014910, 2018.

\bibitem{STAR:2021beb}
M.~S. Abdallah et~al.
\newblock {Global $\Lambda$-hyperon polarization in Au+Au collisions at $\sqrt
  {s_{NN}}$=3~GeV}.
\newblock {\em Phys. Rev. C}, 104(6):L061901, 2021.

\bibitem{ALICE:2019onw}
Shreyasi Acharya et~al.
\newblock {Global polarization of $\Lambda$ and $\bar \Lambda$ hyperons in
  Pb-Pb collisions at $\sqrt {s_{NN}}$ = 2.76 and 5.02 TeV}.
\newblock {\em Phys. Rev. C}, 101(4):044611, 2020.

\bibitem{Kornas:2022cbl}
Fr\'ed\'eric~Julian Kornas.
\newblock {Systematics in the global polarization measurements of $\Lambda$
  hyperons with HADES at SIS18}.
\newblock {\em EPJ Web Conf.}, 259:11016, 2022.

\bibitem{Schilling:1969um}
K.~Schilling, P.~Seyboth, and Guenter~E. Wolf.
\newblock {On the Analysis of Vector Meson Production by Polarized Photons}.
\newblock {\em Nucl. Phys.}, B15:397--412, 1970.
\newblock [Erratum: Nucl. Phys.B18,332(1970)].

\bibitem{Poskanzer:1998yz}
Arthur~M. Poskanzer and S.~A. Voloshin.
\newblock {Methods for analyzing anisotropic flow in relativistic nuclear
  collisions}.
\newblock {\em Phys. Rev. C}, 58:1671--1678, 1998.

\bibitem{Yang:2017sdk}
Yang-Guang Yang, Ren-Hong Fang, Qun Wang, and Xin-Nian Wang.
\newblock {Quark coalescence model for polarized vector mesons and baryons}.
\newblock {\em Phys. Rev. C}, 97(3):034917, 2018.

\bibitem{Xia:2020tyd}
Xiao-Liang Xia, Hui Li, Xu-Guang Huang, and Huan Zhong~Huang.
\newblock {Local spin alignment of vector mesons in relativistic heavy-ion
  collisions}.
\newblock {\em Phys. Lett. B}, 817:136325, 2021.

\bibitem{Gao:2021rom}
Jian-Hua Gao.
\newblock {Helicity polarization in relativistic heavy ion collisions}.
\newblock {\em Phys. Rev. D}, 104(7):076016, 2021.

\bibitem{Becattini:2013vja}
F.~Becattini, L.~Csernai, and D.~J. Wang.
\newblock {$\Lambda$ polarization in peripheral heavy ion collisions}.
\newblock {\em Phys. Rev. C}, 88(3):034905, 2013.
\newblock [Erratum: Phys.Rev.C 93, 069901 (2016)].

\bibitem{Muller:2021hpe}
Berndt M\"uller and Di-Lun Yang.
\newblock {Anomalous spin polarization from turbulent color fields}.
\newblock {\em Phys. Rev. D}, 105(1):L011901, 2022.

\bibitem{Abelev:2008ag}
B.~I. Abelev et~al.
\newblock {Spin alignment measurements of the $K^{*0}$(892) and $\phi$(1020)
  vector mesons in heavy ion collisions at $\sqrt{s_\mathrm {NN}}=200$ GeV}.
\newblock {\em Phys. Rev. C}, 77:061902, 2008.

\bibitem{Anderson:2003ur}
M.~Anderson et~al.
\newblock {The STAR time projection chamber: A Unique tool for studying high
  multiplicity events at RHIC}.
\newblock {\em Nucl. Instrum. Meth.}, A499:659--678, 2003.

\bibitem{Llope:2012zz}
W.~J. Llope.
\newblock {Multigap RPCs in the STAR experiment at RHIC}.
\newblock {\em Nucl. Instrum. Meth.}, A661:S110--S113, 2012.

\bibitem{Acharya:2019vpe}
Shreyasi Acharya et~al.
\newblock {Evidence of Spin-Orbital Angular Momentum Interactions in
  Relativistic Heavy-Ion Collisions}.
\newblock {\em Phys. Rev. Lett.}, 125:012301, 2020.

\bibitem{Serot:1984ey}
Brian~D. Serot and John~Dirk Walecka.
\newblock {The Relativistic Nuclear Many Body Problem}.
\newblock {\em Adv. Nucl. Phys.}, 16:1--327, 1986.

\bibitem{Gasser:1983yg}
J.~Gasser and H.~Leutwyler.
\newblock {Chiral Perturbation Theory to One Loop}.
\newblock {\em Annals Phys.}, 158:142, 1984.

\bibitem{Shifman:1978zn}
Mikhail~A. Shifman, A.~I. Vainshtein, and Valentin~I. Zakharov.
\newblock {Remarks on Higgs Boson Interactions with Nucleons}.
\newblock {\em Phys. Lett. B}, 78:443--446, 1978.

\bibitem{Csernai:2018yok}
L.~P. Csernai, J.~I. Kapusta, and T.~Welle.
\newblock {$\Lambda$ and $\bar{\Lambda}$ spin interaction with meson fields
  generated by the baryon current in high energy nuclear collisions}.
\newblock {\em Phys. Rev. C}, 99(2):021901, 2019.

\bibitem{Karpenko:2016jyx}
I.~Karpenko and F.~Becattini.
\newblock {Study of $\Lambda $ polarization in relativistic nuclear collisions
  at $\sqrt{s_\mathrm {NN}}=7.7$ --200 GeV}.
\newblock {\em Eur. Phys. J. C}, 77(4):213, 2017.

\bibitem{Bryan:1969mp}
R.~Bryan and B.~L. Scott.
\newblock {Nucleon-nucleon scattering from one-boson-exchange potentials. iii.
  s waves included}.
\newblock {\em Phys. Rev.}, 177:1435--1442, 1969.

\bibitem{Nagels:1977ze}
M.~M. Nagels, T.~A. Rijken, and J.~J. de~Swart.
\newblock {A Low-Energy Nucleon-Nucleon Potential from Regge Pole Theory}.
\newblock {\em Phys. Rev. D}, 17:768, 1978.

\bibitem{Walecka:1974qa}
J.~D. Walecka.
\newblock {A Theory of highly condensed matter}.
\newblock {\em Annals Phys.}, 83:491--529, 1974.

\bibitem{Mayer:1949pd}
Mayer Maria~Goeppert.
\newblock {On closed shells in nuclei. II}.
\newblock {\em Phys. Rev.}, 75:1969--1970, 1949.

\bibitem{Haxel:1949fjd}
Otto Haxel, J.~Hans~D. Jensen, and Hans~E. Suess.
\newblock {On the "Magic Numbers" in Nuclear Structure}.
\newblock {\em Phys. Rev.}, 75(11):1766--1766, 1949.

\bibitem{Adler:2000bd}
Clemens Adler, Alexei Denisov, Edmundo Garcia, Michael~J. Murray, Herbert
  Strobele, and Sebastian~N. White.
\newblock {The RHIC zero degree calorimeter}.
\newblock {\em Nucl. Instrum. Meth.}, A470:488--499, 2001.

\bibitem{Whitten:2008zz}
C.~A. Whitten.
\newblock {The beam-beam counter: A local polarimeter at STAR}.
\newblock {\em AIP Conf. Proc.}, 980(1):390--396, 2008.

\bibitem{Allgower:2002zy}
C.~E. Allgower et~al.
\newblock {The STAR endcap electromagnetic calorimeter}.
\newblock {\em Nucl. Instrum. Meth.}, A499:740--750, 2003.

\bibitem{Tang:2018qtu}
A.~H. Tang, B.~Tu, and C.~S. Zhou.
\newblock {Practical considerations for measuring global spin alignment of
  vector mesons in relativistic heavy ion collisions}.
\newblock {\em Phys. Rev. C}, 98(4):044907, 2018.

\bibitem{Abelev:2009bw}
B.I. Abelev et~al.
\newblock {Identified particle production, azimuthal anisotropy, and
  interferometry measurements in Au+Au collisions at $\sqrt{s_{NN}} = 9.2$
  GeV}.
\newblock {\em Phys. Rev. C}, 81:024911, 2010.

\bibitem{Adams:2004ep}
J.~Adams et~al.
\newblock {$K^{*0}$(892) resonance production in Au+Au and p+p collisions at
  $\sqrt{s_{NN}} = 200$ GeV at STAR}.
\newblock {\em Phys. Rev. C}, 71:064902, 2005.

\bibitem{Aggarwal:2010mt}
M.M. Aggarwal et~al.
\newblock {$K^{*0}$ production in Cu+Cu and Au+Au collisions at $\sqrt{s_{NN}}
  = 62.4$ GeV and 200 GeV}.
\newblock {\em Phys. Rev. C}, 84:034909, 2011.

\bibitem{Lan:2017nye}
Shaowei Lan, Zi-Wei Lin, Shusu Shi, and Xu~Sun.
\newblock {Effects of finite coverage on global polarization observables in
  heavy ion collisions}.
\newblock {\em Phys. Lett.}, B780:319--324, 2018.

\bibitem{Fine:2000qx}
V.~Fine and P.~Nevski.
\newblock {OO model of STAR detector for simulation, visualisation and
  reconstruction}.
\newblock In {\em {11th International Conference on Computing in High-Energy
  and Nuclear Physics}}, pages 143--146, 2 2000.

\bibitem{Liang:2020}
Kai-Bao Chen, Zuo-Tang Liang, Yu-Kun Song, and Shu-Yi Wei.
\newblock {Spin alignment of vector mesons in high energy $pp$ collisions}.
\newblock {\em Phys. Rev. D}, 102:034001, 8 2020.

\bibitem{Florkowski:2021pkp}
Wojciech Florkowski and Radoslaw Ryblewski.
\newblock {Interpretation of \ensuremath{\Lambda} spin polarization
  measurements}.
\newblock {\em Phys. Rev. C}, 106(2):024905, 2022.

\end{thebibliography}

\subsection{Data availability}
All raw data for this study were collected using the STAR detector at Brookhaven National Laboratory, and  are not available to the public. Derived data supporting the findings of this study are publicly available in the HEPdata repository (https://www.hepdata.net/record/129067) or from the corresponding author upon request.

\subsection{Code availability}
Codes to process raw data collected by the STAR detector and codes to analyze the produced data are not available to the public.

\subsection{Acknowledgments:}
We thank the RHIC Operations Group and RCF at BNL, the NERSC Center at LBNL, and the Open Science Grid consortium for providing resources and support.  This work was supported in part by the Office of Nuclear Physics within the U.S. DOE Office of Science, the U.S. National Science Foundation, National Natural Science Foundation of China, Chinese Academy of Science, the Ministry of Science and Technology of China and the Chinese Ministry of Education, the Higher Education Sprout Project by Ministry of Education at NCKU, the National Research Foundation of Korea, Czech Science Foundation and Ministry of Education, Youth and Sports of the Czech Republic, Hungarian National Research, Development and Innovation Office, New National Excellency Programme of the Hungarian Ministry of Human Capacities, Department of Atomic Energy and Department of Science and Technology of the Government of India, the National Science Centre of Poland, the Ministry of Science, Education and Sports of the Republic of Croatia, German Bundesministerium f\"ur Bildung, Wissenschaft, Forschung and Technologie (BMBF), Helmholtz Association, Ministry of Education, Culture, Sports, Science, and Technology (MEXT) and Japan Society for the Promotion of Science (JSPS).

\subsection{Author Contribution:}
All authors contributed extensively.

\subsection{Competing interests:}
The authors declare no competing interests.

\subsection{Authors:} 
\author{
M.~S.~Abdallah$^{5}$,
B.~E.~Aboona$^{57}$,
J.~Adam$^{7}$,
L.~Adamczyk$^{2}$,
J.~R.~Adams$^{41}$,
J.~K.~Adkins$^{32}$,
G.~Agakishiev$^{30}$,
I.~Aggarwal$^{43}$,
M.~M.~Aggarwal$^{43}$,
Z.~Ahammed$^{63}$
% ...
}

\normalsize{\rm{(STAR Collaboration)}}

\subsection{Affiliations:}
\normalsize{$^{1}$Abilene Christian University, Abilene, Texas   79699}
\normalsize{$^{2}$AGH University of Science and Technology, FPACS, Cracow 30-059, Poland}
\normalsize{$^{3}$Alikhanov Institute for Theoretical and Experimental Physics NRC "Kurchatov Institute", Moscow 117218}
\normalsize{$^{4}$Argonne National Laboratory, Argonne, Illinois 60439}
\normalsize{$^{5}$American University of Cairo, New Cairo 11835, New Cairo, Egypt}
\normalsize{$^{6}$Ball State University, United States}
\normalsize{$^{7}$Brookhaven National Laboratory, Upton, New York 11973}
\normalsize{$^{8}$University of Calabria \& INFN-Cosenza, Italy}
\normalsize{$^{9}$University of California, Berkeley, California 94720}
\normalsize{$^{10}$University of California, Davis, California 95616}
\normalsize{$^{11}$University of California, Los Angeles, California 90095}
\normalsize{$^{12}$University of California, Riverside, California 92521}
\normalsize{$^{13}$Central China Normal University, Wuhan, Hubei 430079 }
\normalsize{$^{14}$University of Illinois at Chicago, Chicago, Illinois 60607}
\normalsize{$^{15}$Creighton University, Omaha, Nebraska 68178}
\normalsize{$^{16}$Czech Technical University in Prague, FNSPE, Prague 115 19, Czech Republic}
\normalsize{$^{17}$Technische Universit\"at Darmstadt, Darmstadt 64289, Germany}
\normalsize{$^{18}$ELTE E\"otv\"os Lor\'and University, Budapest, Hungary H-1117}
\normalsize{$^{19}$Frankfurt Institute for Advanced Studies FIAS, Frankfurt 60438, Germany}
\normalsize{$^{20}$Fudan University, Shanghai, 200433 }
\normalsize{$^{21}$University of Heidelberg, Heidelberg 69120, Germany }
\normalsize{$^{22}$University of Houston, Houston, Texas 77204}
\normalsize{$^{23}$Huzhou University, Huzhou, Zhejiang  313000}
\normalsize{$^{24}$Indian Institute of Science Education and Research (IISER), Berhampur 760010 , India}
\normalsize{$^{25}$Indian Institute of Science Education and Research (IISER) Tirupati, Tirupati 517507, India}
\normalsize{$^{26}$Indian Institute Technology, Patna, Bihar 801106, India}
\normalsize{$^{27}$Indiana University, Bloomington, Indiana 47408}
\normalsize{$^{28}$Institute of Modern Physics, Chinese Academy of Sciences, Lanzhou, Gansu 730000 }
\normalsize{$^{29}$University of Jammu, Jammu 180001, India}
\normalsize{$^{30}$Joint Institute for Nuclear Research, Dubna 141 980}
\normalsize{$^{31}$Kent State University, Kent, Ohio 44242}
\normalsize{$^{32}$University of Kentucky, Lexington, Kentucky 40506-0055}
\normalsize{$^{33}$Lawrence Berkeley National Laboratory, Berkeley, California 94720}
\normalsize{$^{34}$Lehigh University, Bethlehem, Pennsylvania 18015}
\normalsize{$^{35}$Max-Planck-Institut f\"ur Physik, Munich 80805, Germany}
\normalsize{$^{36}$Michigan State University, East Lansing, Michigan 48824}
\normalsize{$^{37}$National Research Nuclear University MEPhI, Moscow 115409}
\normalsize{$^{38}$National Institute of Science Education and Research, HBNI, Jatni 752050, India}
\normalsize{$^{39}$National Cheng Kung University, Tainan 70101 }
\normalsize{$^{40}$Nuclear Physics Institute of the CAS, Rez 250 68, Czech Republic}
\normalsize{$^{41}$Ohio State University, Columbus, Ohio 43210}
\normalsize{$^{42}$Institute of Nuclear Physics PAN, Cracow 31-342, Poland}
\normalsize{$^{43}$Panjab University, Chandigarh 160014, India}
\normalsize{$^{44}$NRC "Kurchatov Institute", Institute of High Energy Physics, Protvino 142281}
\normalsize{$^{45}$Purdue University, West Lafayette, Indiana 47907}
\normalsize{$^{46}$Rice University, Houston, Texas 77251}
\normalsize{$^{47}$Rutgers University, Piscataway, New Jersey 08854}
\normalsize{$^{48}$Universidade de S\~ao Paulo, S\~ao Paulo, Brazil 05314-970}
\normalsize{$^{49}$University of Science and Technology of China, Hefei, Anhui 230026}
\normalsize{$^{50}$South China Normal University, Guangzhou, Guangdong 510631}
\normalsize{$^{51}$Shandong University, Qingdao, Shandong 266237}
\normalsize{$^{52}$Shanghai Institute of Applied Physics, Chinese Academy of Sciences, Shanghai 201800}
\normalsize{$^{53}$Southern Connecticut State University, New Haven, Connecticut 06515}
\normalsize{$^{54}$State University of New York, Stony Brook, New York 11794}
\normalsize{$^{55}$Instituto de Alta Investigaci\'on, Universidad de Tarapac\'a, Arica 1000000, Chile}
\normalsize{$^{56}$Temple University, Philadelphia, Pennsylvania 19122}
\normalsize{$^{57}$Texas A\&M University, College Station, Texas 77843}
\normalsize{$^{58}$University of Texas, Austin, Texas 78712}
\normalsize{$^{59}$Tsinghua University, Beijing 100084}
\normalsize{$^{60}$University of Tsukuba, Tsukuba, Ibaraki 305-8571, Japan}
\normalsize{$^{61}$United States Naval Academy, Annapolis, Maryland 21402}
\normalsize{$^{62}$Valparaiso University, Valparaiso, Indiana 46383}
\normalsize{$^{63}$Variable Energy Cyclotron Centre, Kolkata 700064, India}
\normalsize{$^{64}$Warsaw University of Technology, Warsaw 00-661, Poland}
\normalsize{$^{65}$Wayne State University, Detroit, Michigan 48201}
\normalsize{$^{66}$Yale University, New Haven, Connecticut 06520}


\clearpage

\section*{Extended Data}

\begin{figure}[h!]       
\centering
\includegraphics[width=0.7\textwidth]{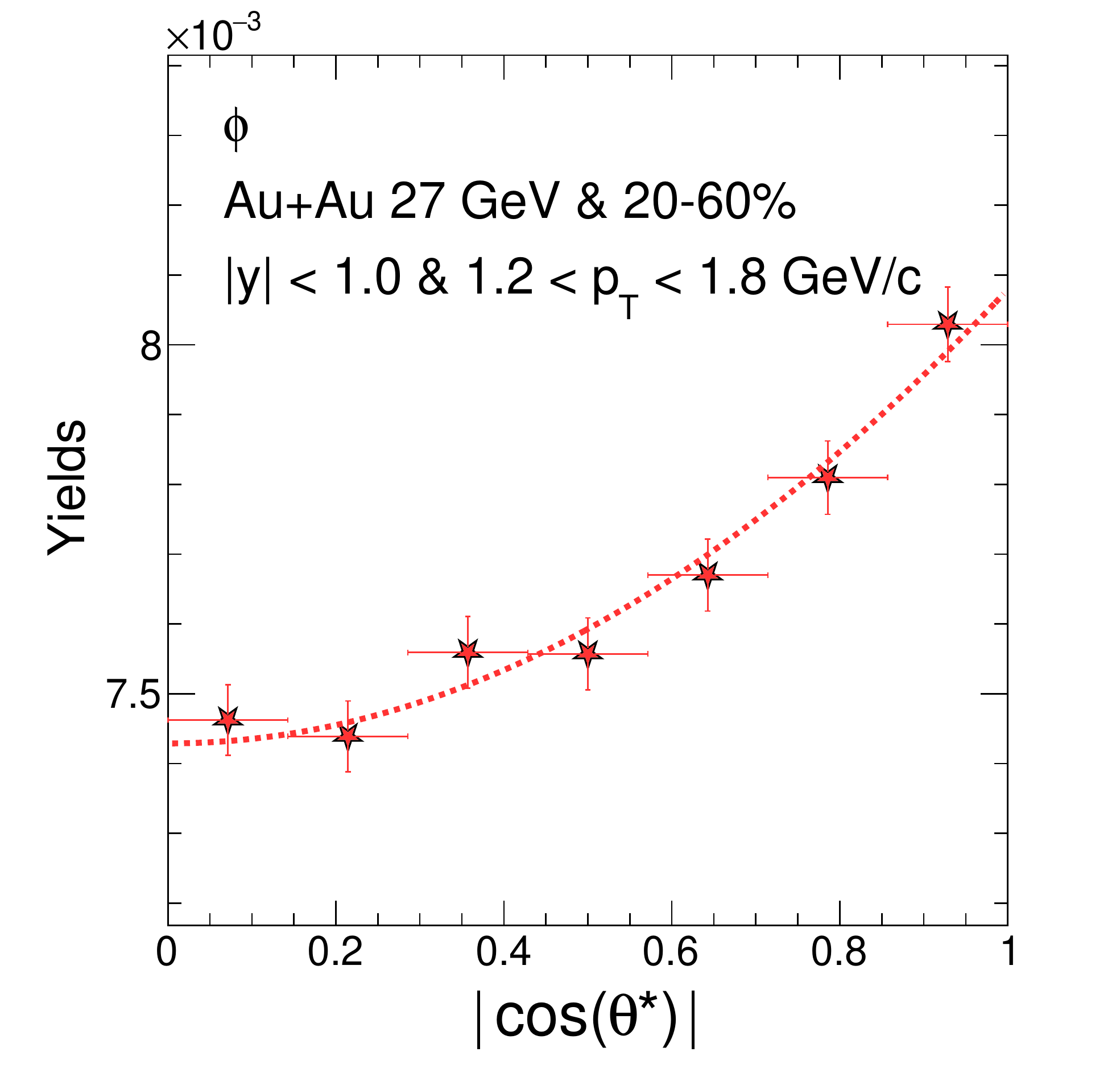}
\caption{\textbf{Efficiency corrected $\phi$-meson yields as a function of cos$\theta$* and corresponding fits with Eq.~\ref{eq:fitEqEffXAccpt_00} in the method section.} The red stars are efficiency corrected yields for $\phi$-mesons with $|y| < 1.0$ and 1.2 $< p_{T} <$ 1.8 GeV/$c$, for 20\%-60\% centrality at $\sqrt{s_{NN}} = 27$ GeV.}
\label{fig:rhoSigCorrPhi}
\end{figure}        
\begin{figure}[h!]       
\centering
\includegraphics[width=0.7\textwidth]{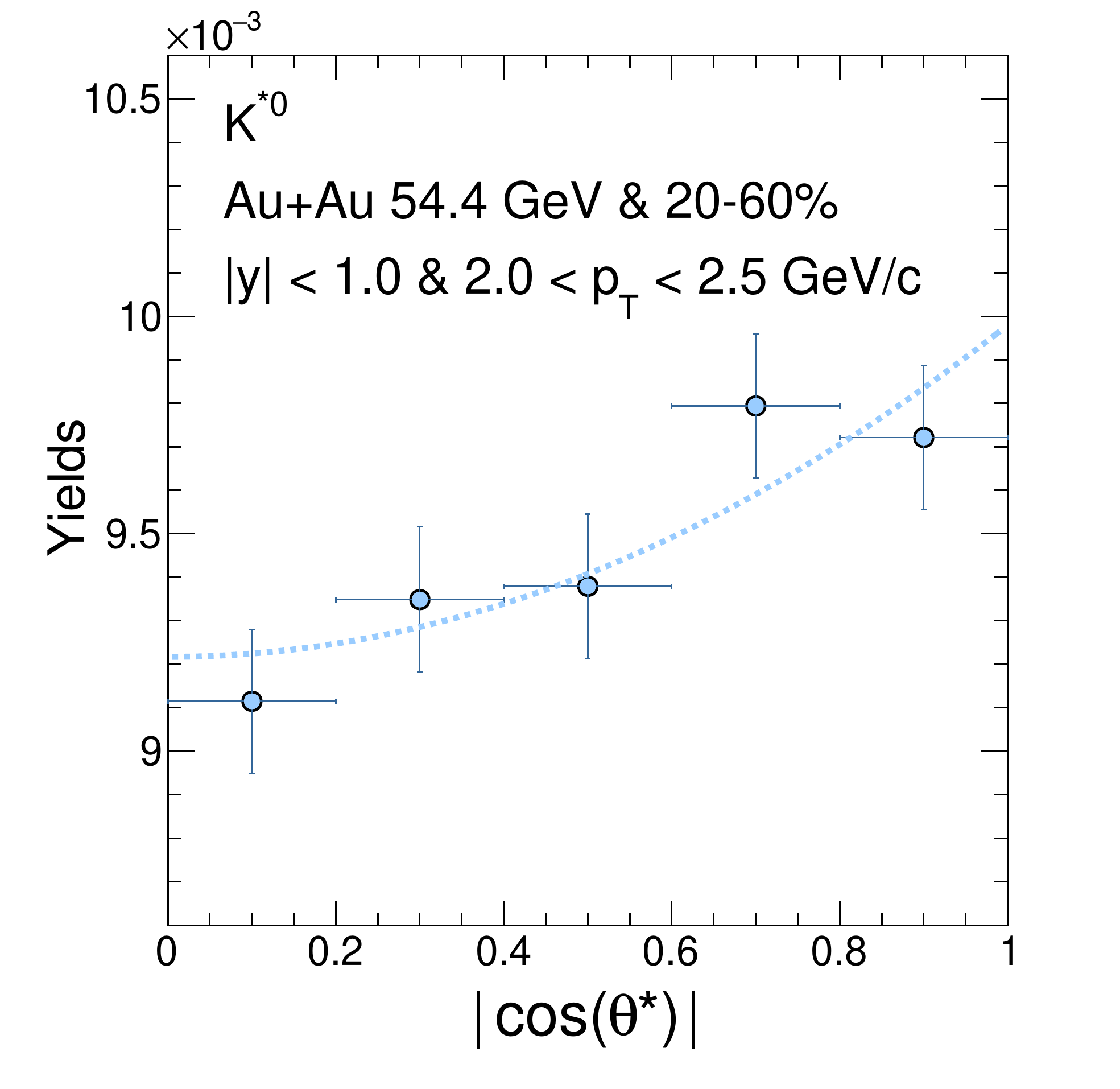}
\caption{\textbf{Efficiency and acceptance corrected $K^{*0}$-meson yields as a function of cos$\theta$* and corresponding fits with Eq.~\ref{eq:abservedRho_00} in the method section.} The blue circles are efficiency and acceptance corrected yields for $K^{*0}$-mesons with $|y| < 1.0$ and 2.0 $< p_{T} <$ 2.5 GeV/$c$, for 20\%-60\% centrality at $\sqrt{s_{NN}} = 54.4$ GeV.}
\label{fig:rhoSigCorrK}
\end{figure}        
\begin{figure}[h!]       
\centering
\includegraphics[width=0.7 \textwidth]{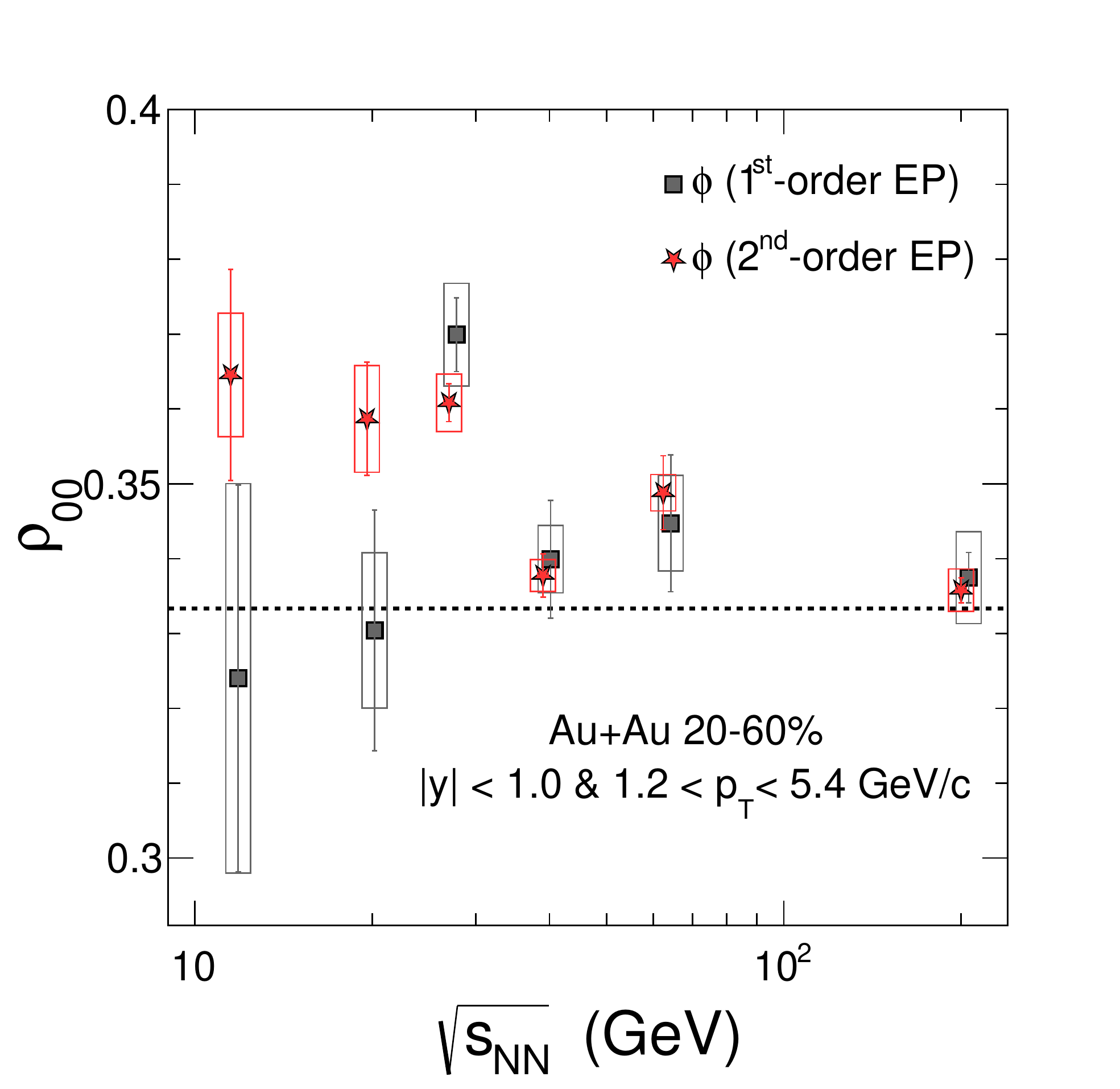}
\caption{\textbf{$\phi$-meson $\rho_{00}$ obtained from 1st- and 2nd-order event planes.} The red stars (gray squares) show the $\phi$-meson $\rho_{00}$ as a function of beam energy, obtained with the 2nd-order (1st-order) EP.}
\label{fig:energyDependence_with1stEP}
\end{figure}        
\begin{figure}[h!]       
\centering
\includegraphics[width=0.7 \textwidth]{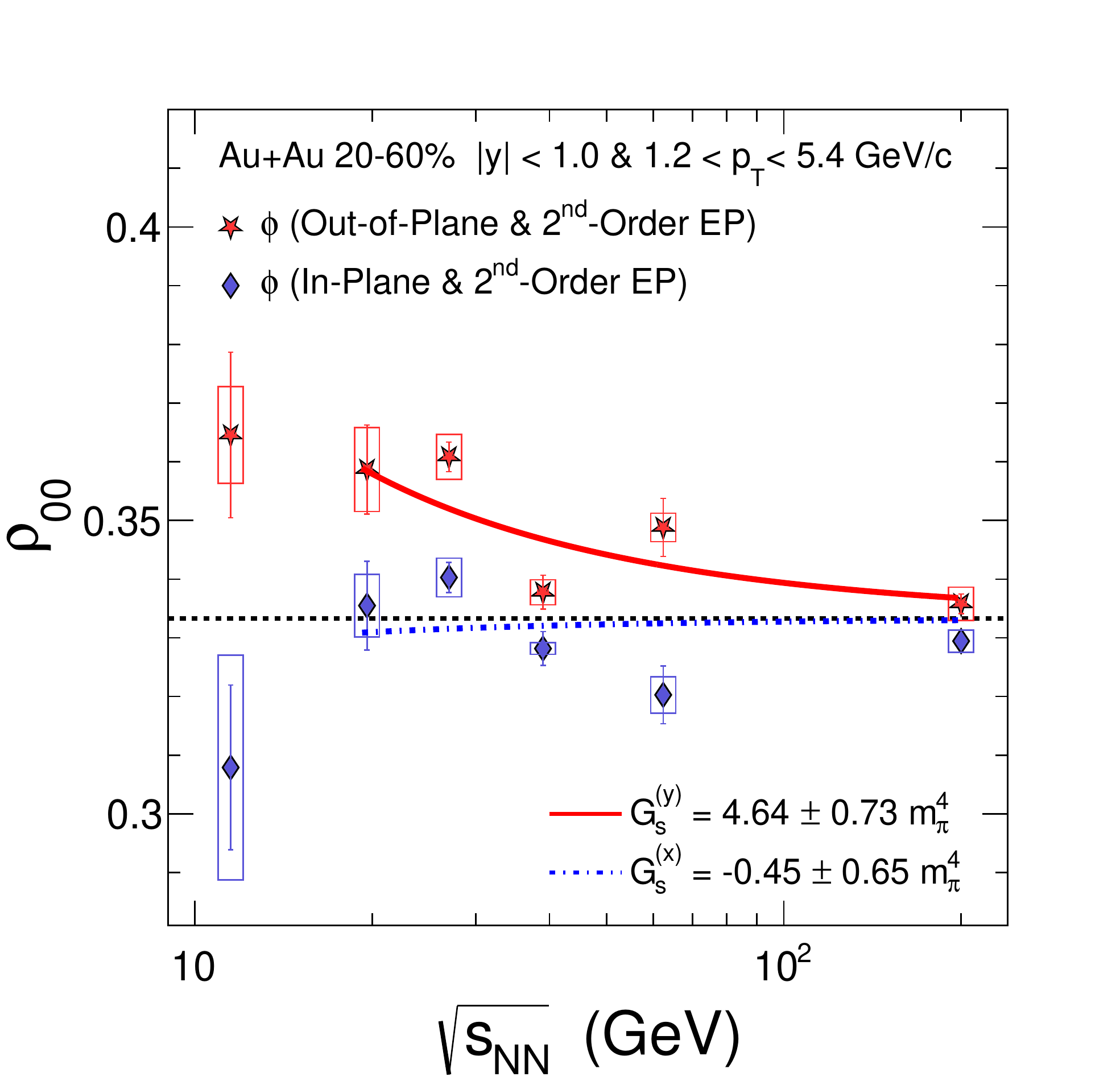}
\caption{\textbf{$\phi$-meson $\rho_{00}$ with respect to different quantization axes.} $\phi$-meson $\rho_{00}$ 
as a function of beam energy, for the out-of-plane direction (stars) and the in-plane direction (diamonds). Curves are fits based on theoretical calculations with a $\phi$-meson field~\cite{Sheng:2019kmk}. The corresponding $G_s$ values obtained from the fits are shown in the legend.}
\label{fig:energyDependence_withInPlane}
\end{figure}
\begin{figure}[h!]       
\centering
\includegraphics[width=0.7 \textwidth]{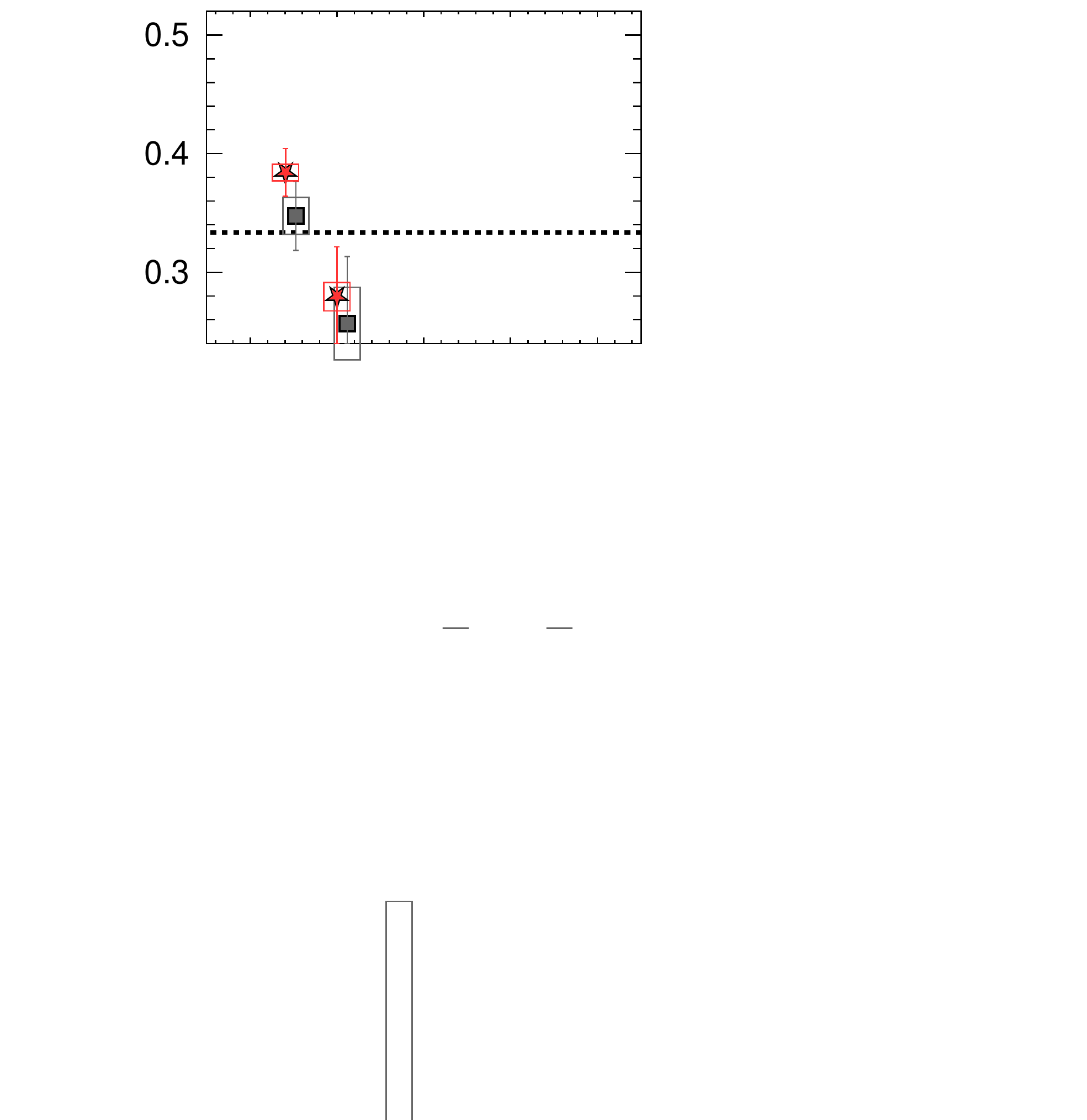}
\caption{\textbf{$\rho_{00}$ as a function of transverse momentum for $\phi$ for different collision energies.}
The gray squares and red stars are results obtained with the 1st- and 2nd-order EP, respectively.
}
\label{fig:ptDependencePhi}
\end{figure}        
\begin{figure}[h!]       
\centering
\includegraphics[width=0.7 \textwidth]{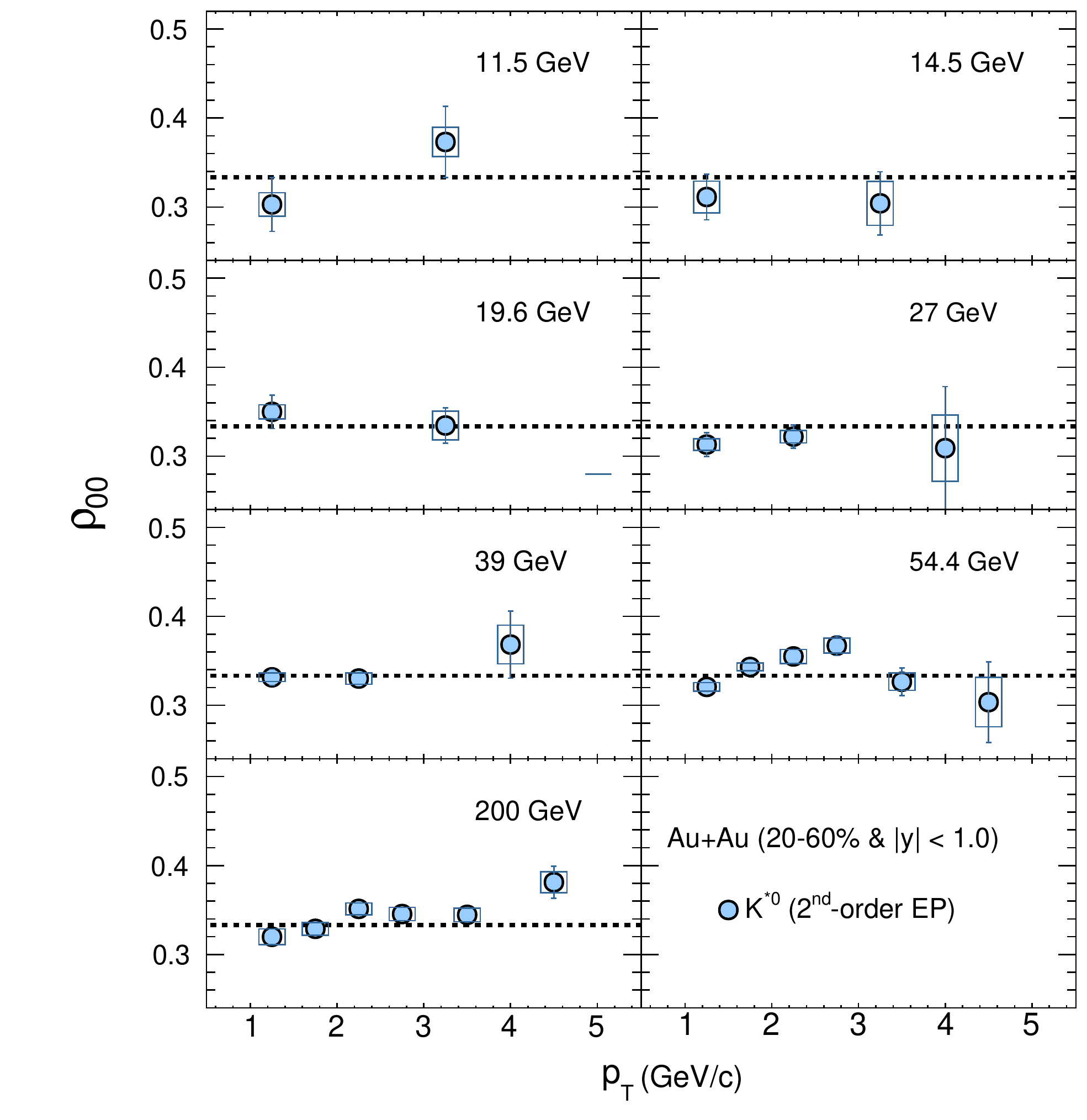}
\caption{\textbf{$\rho_{00}$ as a function of transverse momentum for $K^{*0}$ for different collision energies.} The solid circles are results obtained with the 2nd-order EP.}
\label{fig:ptDependenceKstar}
\end{figure}        
\begin{figure}[h!]       
\centering
\includegraphics[width=0.8 \textwidth]{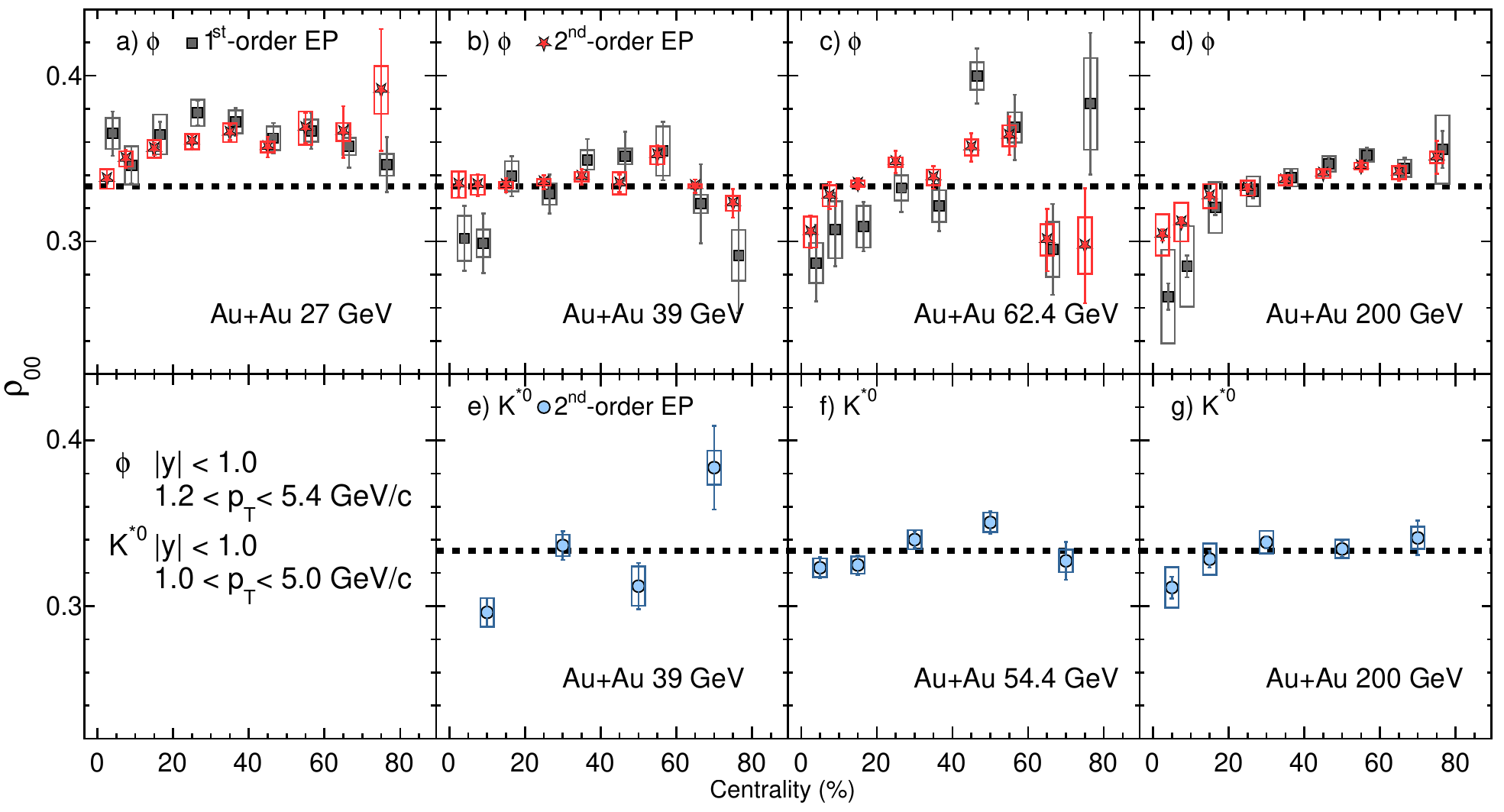}
\caption{\textbf{$\rho_{00}$ as a function of centrality for $\phi$ (upper panels) and $K^{*0}$ (lower panels).} The solid squares and stars are results for the $\phi$ meson, obtained with the 1st- and 2nd-order EP, respectively. The solid circles are results for the $K^{*0}$ meson, obtained with the 2nd-order EP.}
\label{fig:centDependence}
\end{figure}        
\begin{figure}[h!]       
\centering
\includegraphics[width=0.8 \textwidth]{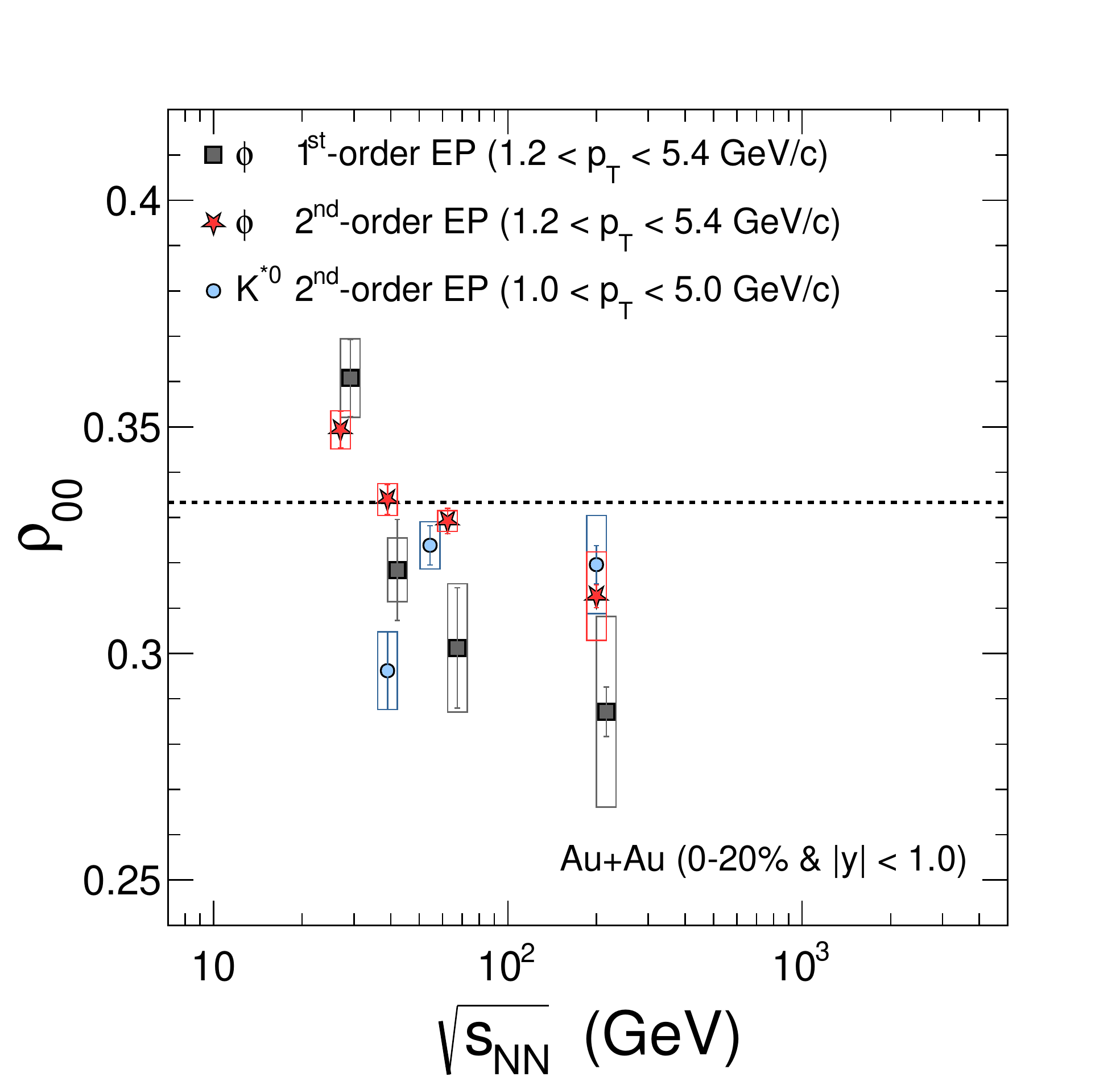}
\caption{\textbf{Global spin alignment measurement of $\phi$ and $K^{*0}$ vector mesons in Au+Au collisions at 0-20\% centrality.} The solid squares and stars are results for the $\phi$ meson, obtained with the 1st- and 2nd-order EP, respectively. The solid circles are results for $K^{*0}$-meson, obtained with the 2nd-order EP.}
\label{fig:energyDependenceCentral}
\end{figure}        

\clearpage

\begin{table}[htb!]\centering
\caption{Sources of systematic error in $\phi$. The tabulated numbers are absolute uncertainties in $\rho_{00}$.}
  \begin{tabular}{c c c c c c}
  \hline
    & Quality Cuts & PID Cuts & Signal \& Yields Extraction & Efficiency & Total \\
  \hline
  1st-order EP & 0.0015 & 0.0017 & 0.0031 & 0.0017 & 0.0042 \\
  \hline
  2nd-order EP & 0.0006 & 0.0006 & 0.0013 & 0.0005 & 0.0017  \\
  \hline 
  \end{tabular}
\label{table:phiSysError}
\end{table}

\begin{table}[htb!]\centering
\caption{Sources of systematic error in $K^{*0}$. The tabulated numbers are absolute uncertainties in $\rho_{00}$.}
  \begin{tabular}{c c c c c c}
  \hline
    & Quality Cuts & PID Cuts & Signal Extraction & Yields Extraction & Total \\
  \hline
  2nd-order EP & 0.0018 & 0.0020 & 0.0030 & 0.0015 & 0.0043 \\
  \hline
  \end{tabular}
\label{table:kStarSysError}
\end{table}

\clearpage

\end{document}